\documentclass[a4paper,twocolumn,11pt,accepted=2024-05-01]{quantumarticle}
\pdfoutput=1
\usepackage[utf8]{inputenc}
\usepackage[english]{babel}
\usepackage[T1]{fontenc}
\usepackage{amsmath}
\usepackage{amsthm}
\usepackage{amssymb}
\usepackage{bbold}
\usepackage{hyperref}
\usepackage{csquotes}
\usepackage{booktabs}
\usepackage[style=base]{caption}

\usepackage{cite}
\usepackage{tikz}
\usepackage{lipsum}

\begin{document}

\title{Precisely determining photon-number in real time}

\author{Leonardo Assis Morais}
\affiliation{Centre for Engineered Quantum Systems, University of Queensland, QLD 4072, Australia}
\affiliation{School of Mathematics and Physics, University of Queensland, QLD 4072, Australia}
\orcid{0000-0003-1617-979X}
\email{leoassisfisica@gmail.com}
\author{Till Weinhold}
\affiliation{Centre for Engineered Quantum Systems, University of Queensland, QLD 4072, Australia}
\affiliation{School of Mathematics and Physics, University of Queensland, QLD 4072, Australia}
\author{Marcelo Pereira de Almeida}
\affiliation{Centre for Engineered Quantum Systems, University of Queensland, QLD 4072, Australia}
\affiliation{School of Mathematics and Physics, University of Queensland, QLD 4072, Australia}
\author{Joshua Combes}
\affiliation{Department of Electrical, Computer and Energy Engineering, University of Colorado Boulder, Boulder, Colorado 80309, USA}
\author{Markus Rambach}
\affiliation{Centre for Engineered Quantum Systems, University of Queensland, QLD 4072, Australia}
\affiliation{School of Mathematics and Physics, University of Queensland, QLD 4072, Australia}
\orcid{0000-0002-4659-3804}
\author{Adriana Lita}
\affiliation{National Institute of Standards and Technology, 325 Broadway, Boulder, Colorado 80305, USA}
\author{Thomas Gerrits}
\affiliation{National Institute of Standards and Technology, 325 Broadway, Boulder, Colorado 80305, USA}
\author{Sae Woo Nam}
\affiliation{National Institute of Standards and Technology, 325 Broadway, Boulder, Colorado 80305, USA}
\author{Andrew G. White}
\affiliation{Centre for Engineered Quantum Systems, University of Queensland, QLD 4072, Australia}
\affiliation{School of Mathematics and Physics, University of Queensland, QLD 4072, Australia}
\author{Geoff Gillett}
\affiliation{Centre for Engineered Quantum Systems, University of Queensland, QLD 4072, Australia}
\affiliation{School of Mathematics and Physics,  University of Queensland, QLD 4072, Australia}
\affiliation{Quantum Valley Ideas Lab, Waterloo, ON N2L 6R2, Canada}

\maketitle
\begin{abstract}
Superconducting transition-edge sensors (TES) are sensitive microcalorimeters used as photon detectors with unparalleled energy resolution. They have found applications from measuring astronomical spectra through to determining the quantum property of photon-number, $\hat{n} {=}\hat{a}^{\dag}\hat{a}$, for energies from 0.6--2.33~eV. However, achieving optimal energy resolution requires considerable data acquisition---on the order of 1~GB/min---followed by post-processing, preventing real-time access to energy information. We report a custom hardware processor to process TES pulses while new detection events are still being registered, allowing the photon-number to be measured in real time. We resolve photon-number up to $n{=}16$---achieving up to parts-per-billion discrimination for low photon-numbers on the fly---providing transformational capacity for TES applications from astronomy to quantum technology.
\end{abstract}

\section{Introduction}
\label{sec:intro}

Superconducting transition-edge sensors (TES) are currently used in a wide range of disciplines, with applications ranging from astronomy, where they are used to count photons at the output of a spectrometer~\cite{romani_first_1999, burney_transition-edge_2006, smith_multiabsorber_2019}, through to quantum photonics, where output photons are counted in applications that have a known source spectrum~\cite{obrien_photonic_2009, hadfield_single-photon_2009, slussarenko_photonic_2019, quesada_simulating_2019, su_conversion_2019, tzitrin_progress_2020}. 
TES are single-photon detectors with a set of exquisite properties. These detectors achieve energy resolution of 0.15~eV from the infrared to ultraviolet spectral range~\cite{cabrera_detection_1998, miller_demonstration_2003}; possess an exceedingly low dark count rate---even at low photon flux~\cite{rosenberg_noise-free_2005}; and operate with high efficiency---up to 98\%~\cite{lita_counting_2008, fukuda_titanium-based_2011}. These characteristics combined make TES a photon detector with unique photon-number-resolving (PNR) attributes compared to other platforms like superconducting nanowire single-photon detectors (SNSPDs). 

SNSPDs working principle differs fundamentally from that of TES, allowing them to achieve high detection efficiencies with fast detection rates, but they cannot intrinsically discriminate larger photon numbers~\cite{stasi-PSNSPD_2023}.
Different routes are being explored towards number resolution such as pulse analysis~\cite{cahall_multi-photon_2017} or parallelisation~\cite{stasi-PSNSPD_2023,tao_high_2019, mattioli_photon-counting_2016}.
Nevertheless, both approaches have significant limitations regarding PNR and efficiency and pose significant challenges to electronic readout
(see Table~\ref{tab:comparison} Appendix~\ref{appendix:comparison} for a comparison of PNR implementations).

A central challenge for TES is their slower recovery ($1/e$ decay) time, typically on the order of $\mu s$.
Recent developments have improved this to less than 250~ns~\cite{hummatov_fast_2023}, allowing photon detection rates in the MHz range.
For TES there exists a fundamental trade-off between achievable rates and PNR, as faster detectors either require higher operation temperatures and reduced detector area~\cite{portesi_fabrication_2015} or additional gold heat sinks~\cite{kobayashi_development_2019, hummatov_fast_2023}, with both methods degrading the energy resolution.
However parallelisation similar to SNSPDs~\cite{eaton_resolution_2023} has been demonstrated as a promising path forward with less trade-offs.

Typically, energy information---be it spectrum or photon-number depending on the implementation---is extracted from a TES by measuring the time varying voltage signal produced by each detection event. 
Usually, the entire data stream from the TES detector output is recorded and post-processed in software~\cite{levine_algorithm_2012, humphreys_tomography_2015}. 
Here, we introduce a different approach: a reconfigurable hardware processor, implemented with a Field-Programmable Gate Array (FPGA), that measures and stores a handful of physically significant pulse properties for every pulse,  before discarding the voltage signal.

We have implemented a flexible and powerful platform to determine photon-number. Its capabilities include: digitisation of the TES analogue output; identification of pulses related to photon detection events in the digitised data stream; time stamping of detection events; measurement of relevant pulse characteristics to perform photon-number-assignment tasks; and, optionally, recording of the entire pulse, for calibration or cases where optimal information about detection events is required.

\section{TES Signal and Analysis}
\label{sec:tes_signal}

\begin{figure}[!t]
    \includegraphics[width=0.49\textwidth]{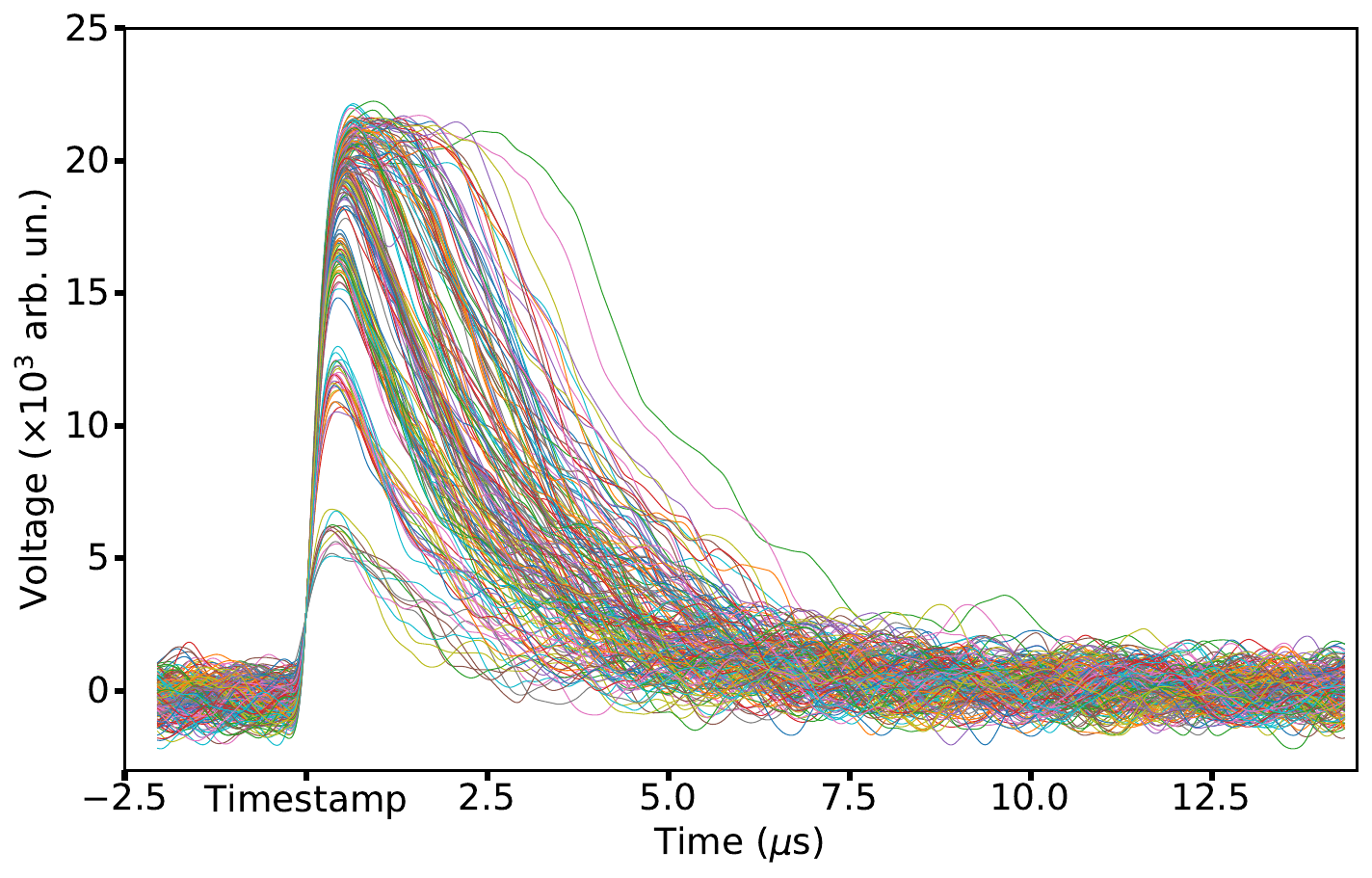}
    \includegraphics[width=0.49\textwidth]{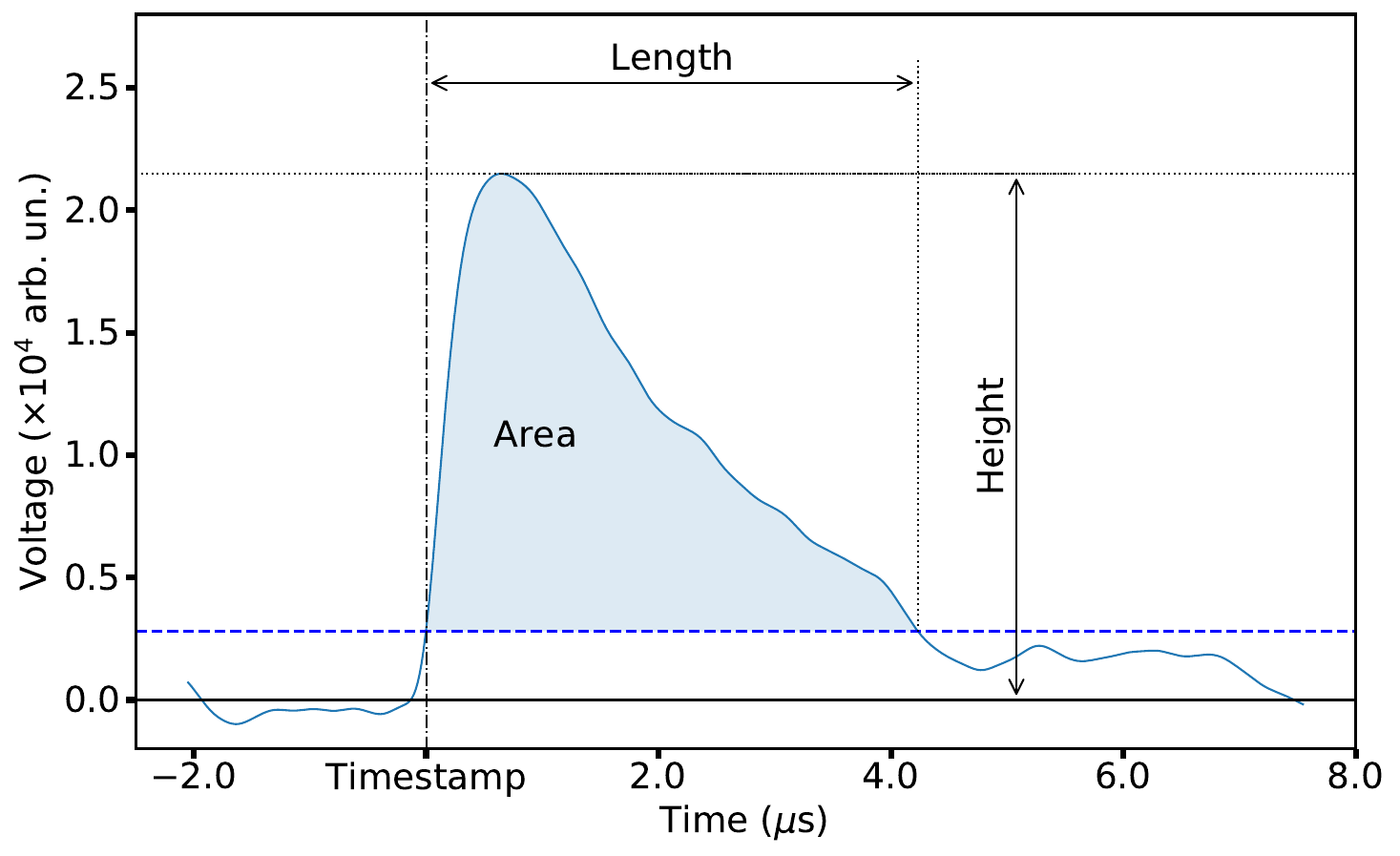}
    \vspace{-7mm}
    \caption{(\textit{Top}) 200 TES pulses captured by our hardware processor using 820~nm pulsed light from a strongly attenuated diode laser. The voltage reading is proportional to the current change due to a photon wavepacket absorption. By keeping the sensor in the transition region between its superconducting and normal phases, we can measure energy with enough precision to distinguish the pulse associated with different photon numbers. 
    (\textit{Bottom}) Single TES pulse digitised by our FPGA showing some of the measurements our system can deliver. The solid black line is the detection baseline, and the dash-dotted blue line is the height detection threshold. \vspace{-3mm}}
   \label{fig:traces_and_measures}
\end{figure}

The top of Figure~\ref{fig:traces_and_measures} shows 200 detection events for a weak coherent pulse at 820~nm, registered using our hardware processor. After a detection is performed, a sharp rise is observed in the TES output, followed by an exponential-like decay whose constant will depend on the thin film properties~\cite{calkins_faster_2011, lolli_characterization_2013}. The sharp rises are of the order of hundreds of nanoseconds, and the decay time varies from 2 to 8~$\mu$s, depending on the energy of the absorbed photon wavepacket. The higher the number of photons in a detection event, the longer it takes for the detector to cool down to its operating temperature $T_0$.

The bottom of Figure~\ref{fig:traces_and_measures} highlights the input parameters used to register a valid detection event and the key characteristics subsequently recorded (see Appendix~\ref{appendix:fpga} for additional characteristics). The user sets: a detection baseline (solid black line) as the reference level with respect to the height measurements; the height threshold (blue-dashed line); and the slope threshold (see Figure~\ref{fig:fpga_with_all} in Appendix~\ref{appendix:fpga} for further details). To be considered a valid detection event, a TES pulse must have both height and slope higher than their respective thresholds. Only when this condition is met are the event packets---structures that gather all the information about the measurements---created and recorded. 

\begin{figure*}[!t]
    \centering
    \includegraphics[width=0.325\textwidth]{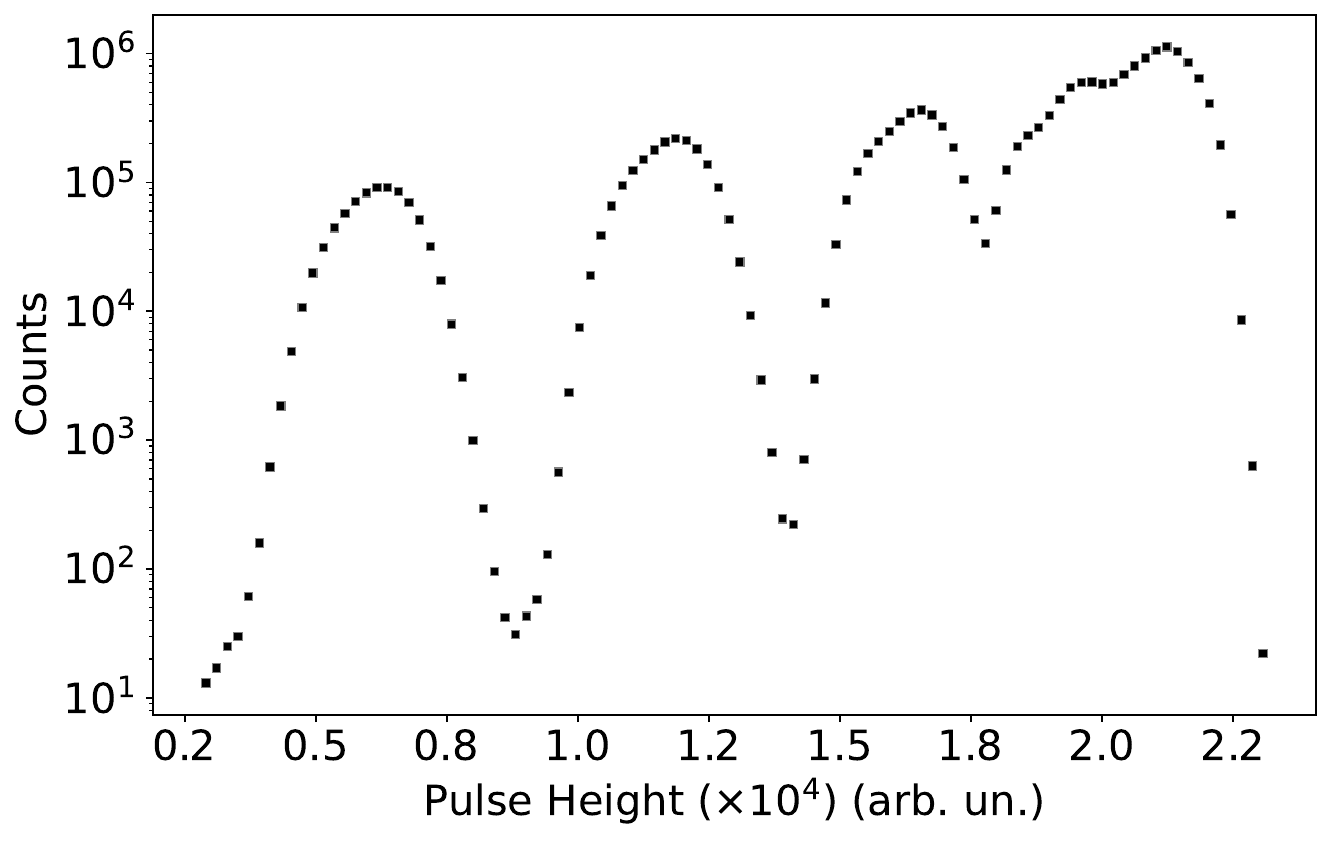}
    \includegraphics[width=0.325\textwidth]{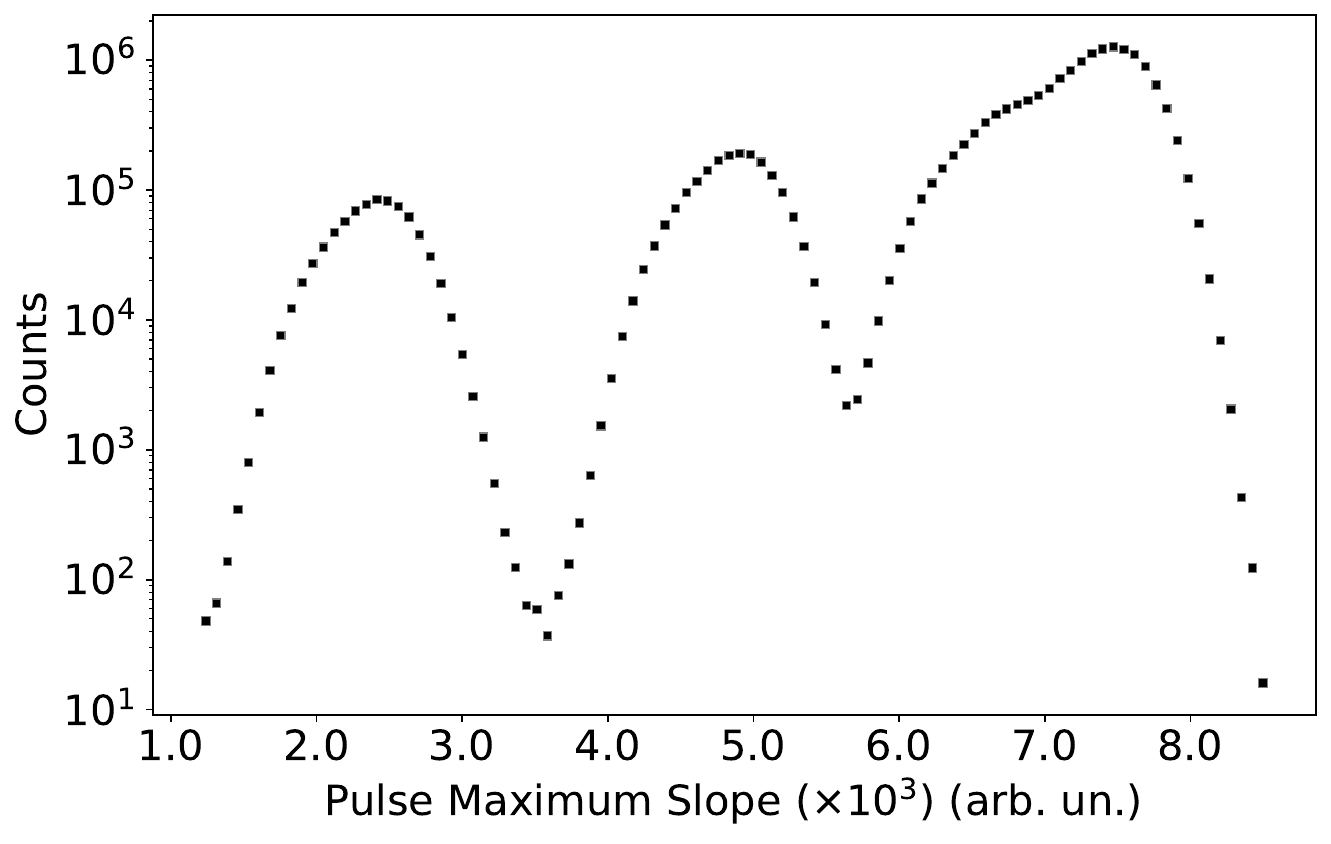}
     \includegraphics[width=0.325\textwidth]{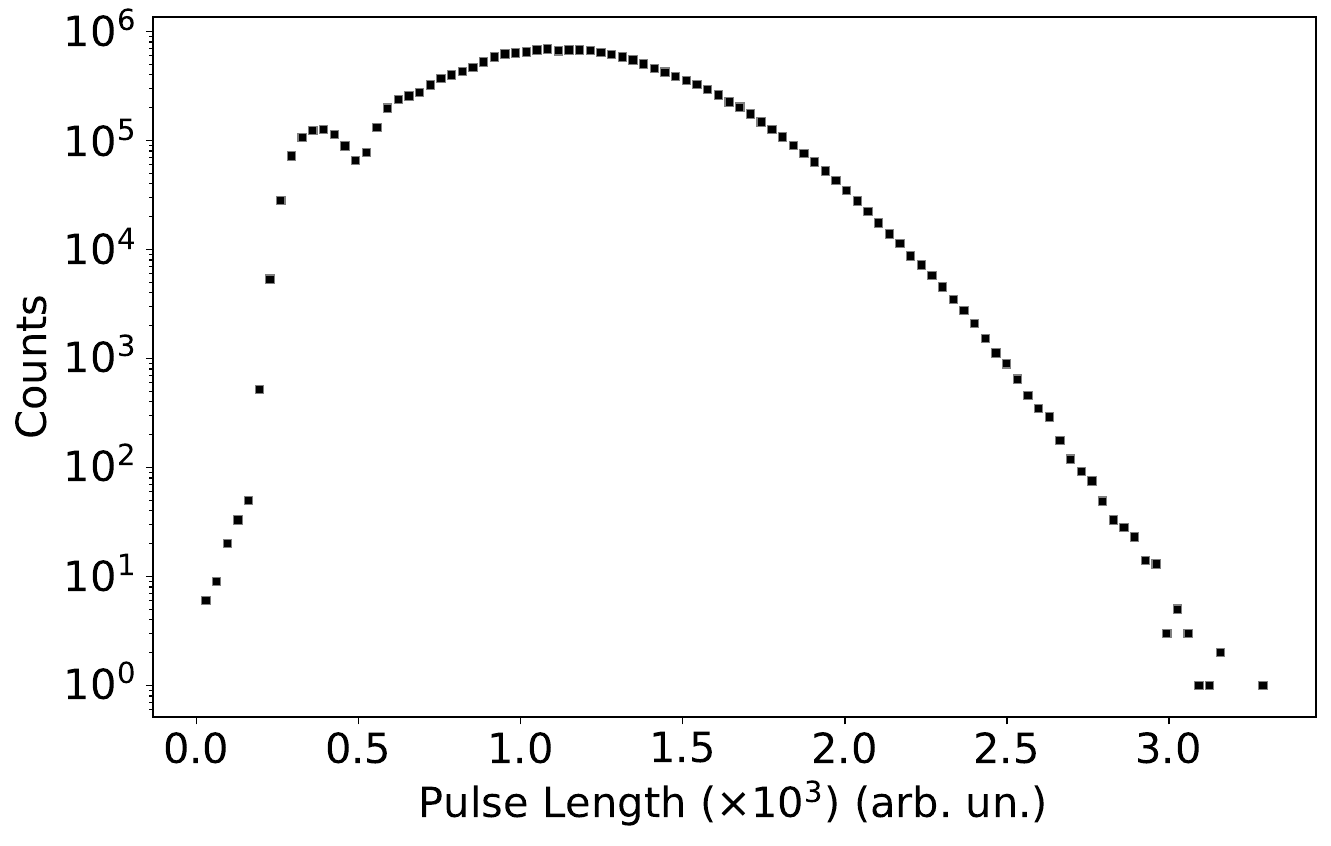}
    \caption{Histograms for photon detection events using a strongly attenuated pulsed diode laser at 820~nm for different aspects of the TES pulse. (\textit{Left}) Height. Up to three peaks are clearly resolved. For a higher number of photons, the detector transitions to the normal phase, making it impossible to distinguish the photon number using height. (\textit{Middle}) Maximum slope. Only 2 peaks are clearly resolved. We believe that the bandwidth filters used to remove high frequency noise compromise our ability to use pulse slope to distinguish photon number. (\textit{Right}) Length. Due to signal noise, there is poor differentiation between events with one or more photons. \vspace{-3mm}}
    \label{fig:histogram_characteristics}
\end{figure*}

To assist the configuration of the baseline and detection thresholds, our processor contains a built-in multi-channel analyser (MCA). The MCA broadcasts histograms of a single quantity of the digitised output from the TES detectors in real time. We use the MCA to analyse the noise levels appropriately and set the baseline and detection thresholds. To determine the position of the detection baseline, we monitor the noise distribution, i.e. the detector output when no light is sent to the detectors. The detection baseline is chosen to be on the mode of the noise distribution. To determine the detection thresholds, we use the same light source which will be used in the experiment. 
These thresholds are positioned between the end of the noise distribution and the first peak related to actual photon detections. Since anything lower than these thresholds will not be considered a valid photon detection event, these parameters will set the effective efficiency of the performed detections. Their appropriate values will depend on the experiment conditions, and a trade-off between detection efficiency and background counts must be considered when finding their optimal values. 

After setting the baseline and detection thresholds, our system is ready to perform measurements over different characteristics of the TES pulse. By storing only the pulse characteristics instead of the entire pulse, we achieve a 40 times reduction in transferred and stored data. The measured values are available in the FPGA less than 500~ns after the pulse falls below the height threshold. An Ethernet connection transfers pulse data to the computer, where photon number information is obtained within milliseconds after registering a pulse event. As TES are limited by their response time to the kHz counting regime, our photon number identification is fast enough to detect events in real time.

\section{Detector Calibration}
\label{sec:results}

\begin{figure}[!t]
    \includegraphics[width=\columnwidth]{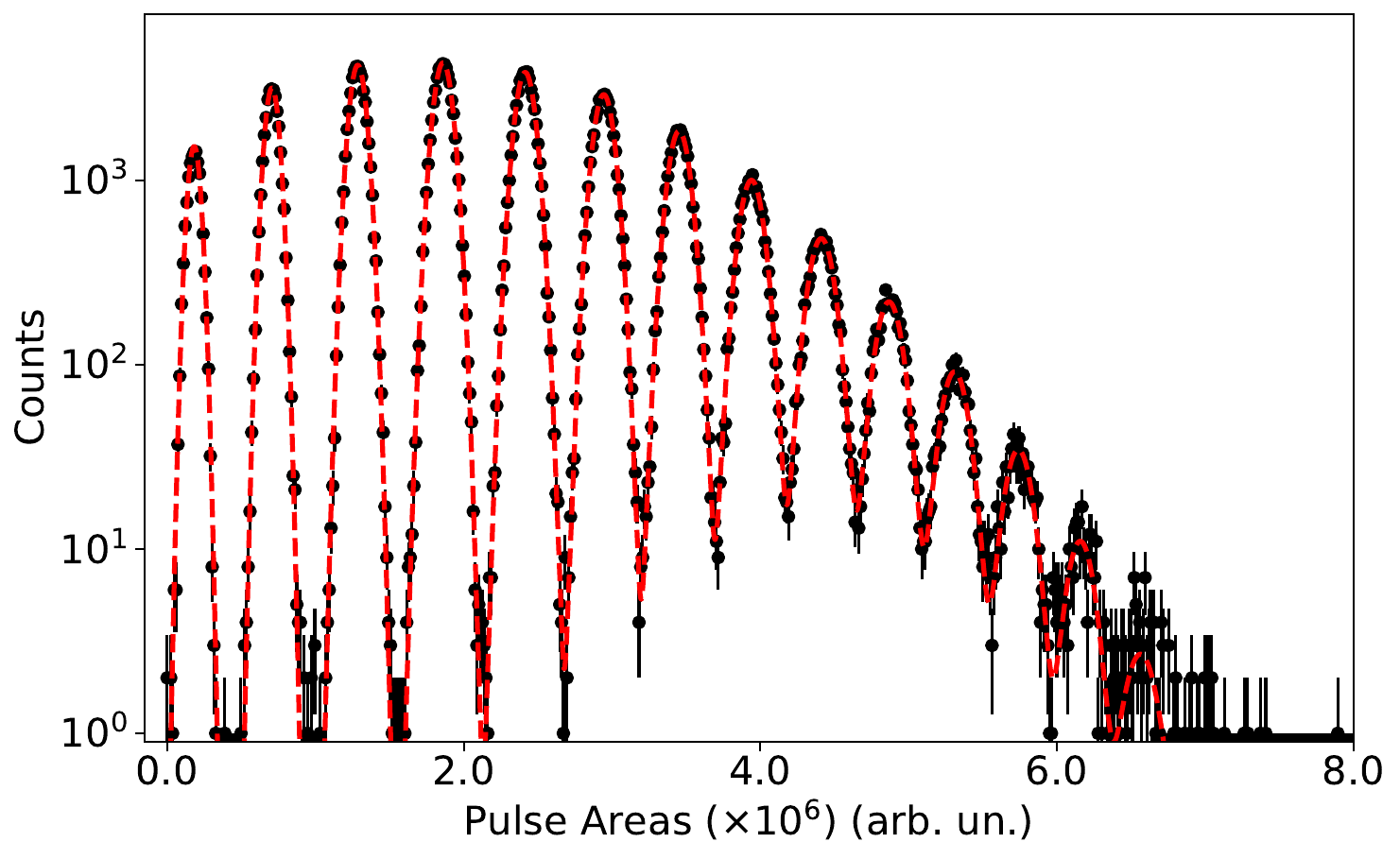}
     \vspace{-7mm}
    \caption{TES Pulse area histogram (black circles) for a strongly attenuated pulsed diode laser at 820~nm. Dashed red line is the model comprised of a sum of 16 Gaussian curves (see Appendix~\ref{appendix:area_histogram} for details). There are no peaks associated with zero photons due to the triggering requirement, as explained in the text. Note that the last two Gaussian curves are not visible due to their low amplitude.  \vspace{-3mm}}
    \label{fig:histogram_area}
\end{figure}

To precisely determine the best way to assign a photon number to a detection event, we measured and analysed four TES pulse characteristics: area, height, length, and maximum slope, as shown in the bottom of Figure~\ref{fig:traces_and_measures}. \textit{Height} is measured as the distance between the baseline and the maximum point in the TES pulse. \textit{Length} is the distance between the rising pulse crossing the detection threshold and the next point where the decaying pulse crosses this threshold again. \textit{Area} is calculated by summing the heights between the detection threshold and the pulse every 4~ns, as shown in the shaded region. \textit{Slope} is determined by differentiating the pulse, with the maximum slope being the highest value achieved by the slope signal (see Figure~\ref{fig:fpga_with_all} in Appendix~\ref{appendix:fpga}). We show typical histograms for these quantities in Figures~\ref{fig:histogram_characteristics} and \ref{fig:histogram_area}. We found that the area provides, by far, the best discrimination (as seen in the figures), as well as the highest precision. This reflects the fact that for TES, the pulse area is proportional to the energy absorbed during the detection event---even after the TES undergo the transition to the normal phase occurs~\cite{irwin_application_1995}---and TES takes longer to cool down after absorbing a photon wavepacket with higher photon number (see top in Figure \ref{fig:traces_and_measures}).

\begin{figure*}[t]
  \begin{minipage}{1.2\columnwidth}
    {\includegraphics[width=\columnwidth]{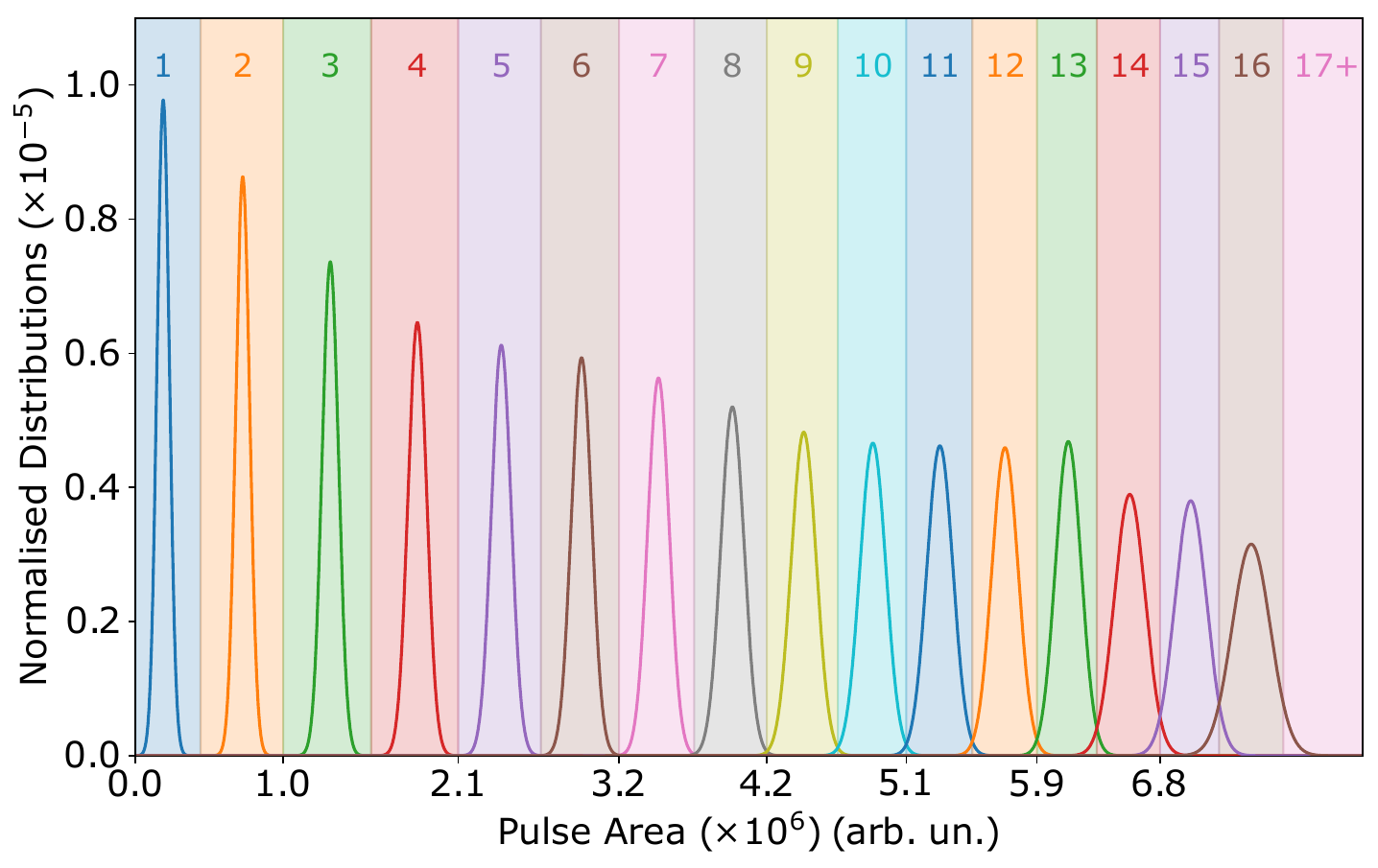}}
  \end{minipage}
    \begin{minipage}{0.8\columnwidth}
    {\includegraphics[width=\columnwidth]{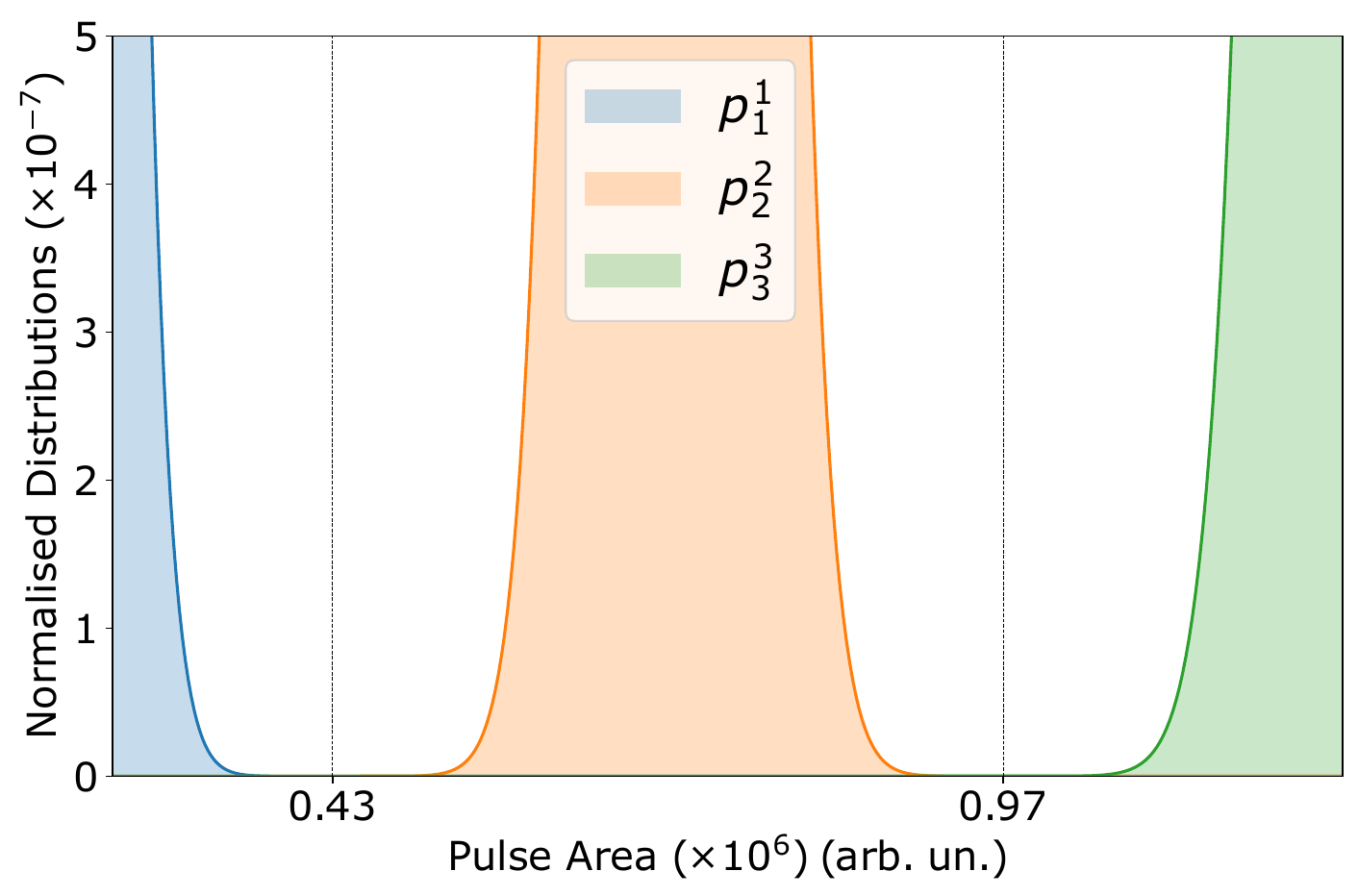}}
    {\includegraphics[width=\columnwidth]{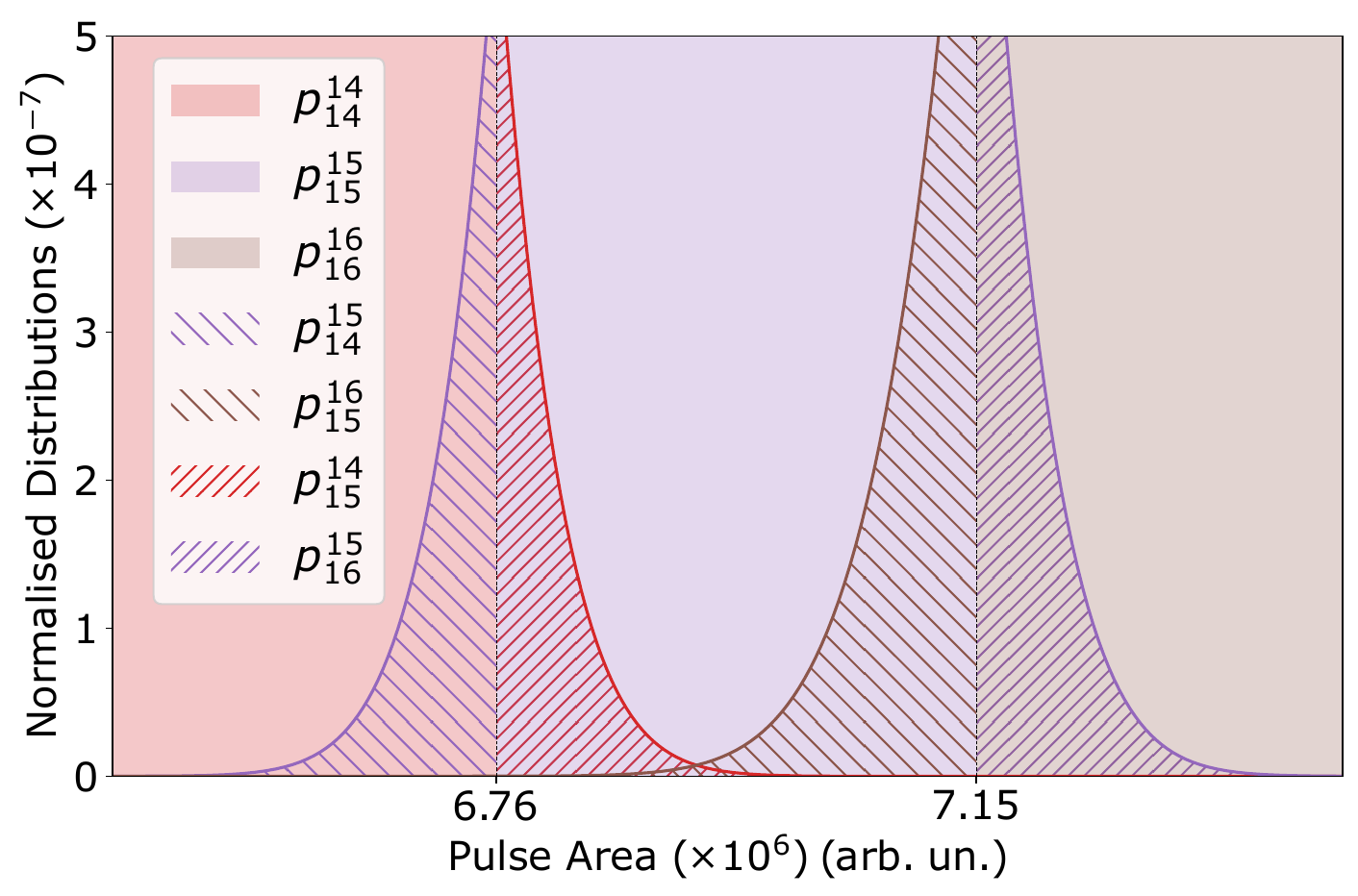}}
  \end{minipage}
  \vspace{-4mm}
\caption{(\textit{Left}) Solid curves are normalised area distributions, $g_n(a)$. The counting thresholds $t_m$ are positioned in the intersection between adjacent distributions. The probability of assigning $m$ photons to a $n$ photon detection event $p^n_m$ is the percentage of area of $g_n(a)$ within a coloured interval $[t_m, t_{m+1}]$. (\textit{Right}) Insets highlighting the overlaps between three adjacent distributions. (\textit{Top Right}) Centred at $n {=} 2$, we see $p^2_2 {=} 0.99999999527^{(2)}_{(2)}$ (number in parenthesis corresponds to one standard deviation), and $p^2_1 {=} 8.5 {\times} 10^{-10} $, $p^2_3 {=} 3.88 {\times} 10^{-9}$. The overlaps between these distributions are too small to be seen in the inset. (\textit{Bottom Right})  Centred at $n {=} 15$, we see $p^{15}_{15} {=} 0.936^{(4)}_{(7)} $, and $p^{15}_{14} = 2.6 {\times} 10^{-2}$, $p^{15}_{16} {=} 3.7 {\times} 10^{-2}$.\vspace{-3mm}}
 \label{fig:normalised}
\end{figure*}

The important question is: given that a detection event is recorded, what is the chance of assigning a photon number $m$ to a detection event with a true number of $n$ photons? This probability is called discrimination. To evaluate it, we tested our system using a strongly attenuated pulsed diode laser with emission centred at 820~nm, described by the weak coherent state:
\begin{equation}
   |\alpha \rangle = e^{|\alpha|^2/2} \sum_{n=0}^{\infty} \frac{\alpha^n}{(n!)^{1/2}} |n \rangle,
\end{equation}
\noindent where $|\alpha|^2$ is proportional to the intensity of the beam. 

A series of neutral density filters was employed to attenuate the optical power of our light source. Fine light intensity control was achieved by employing a pair of linear polarisers. To guarantee that our source was quasi-monochromatic, we used an 820~nm narrow-band filter with 2~nm bandwidth. The light was coupled to a single mode fibre which was directly connected to the TES detectors (see Appendix~\ref{appendix:tes-detectors} for more details). Our optical setup was optically isolated to prevent noise contamination. To avoid multi-absorption events (as shown e.g. in Figure~\ref{fig:multi_absorption}, Appendix~\ref{appendix:multi-events}), we used a pulsed light source. The laser was driven with electrical pulses of 2.6~V amplitude and 50~ns width, using a repetition rate of 10~kHz. This guarantees a time separation of at least 100~$\mu$s between subsequent detection events. The TES pulse was amplified at room temperature using a pre-amplifier and differential amplifier, which also applied a 1~MHz bandwidth filter to remove high frequency noise. The filtered signal was digitised by a 14-bit analogue-to-digital converter at a rate of 250~MHz. The digitised signal was the input for our FPGA hardware processor. Our current implementation uses up to two channels simultaneously (see Appendix~\ref{appendix:fpga} for additional details on the FPGA). For further information about the hardware processor implementation, see Ref.~\cite{gillett_hardware_2018}. We measured a background count rate of $0.26 \pm 0.04$~counts/s.

A typical histogram of TES pulse areas is shown in Figure~\ref{fig:histogram_area} (see Figure~\ref{fig:intensity_diff} Appendix~\ref{appendix:area_histogram} for histograms with different intensities). 
We determined the number of bins in the histogram by analysing how the reduced chi-square $\chi^2_R $ varies with the bin number (see Appendix~\ref{appendix:area_histogram}). 
Each peak on the TES pulse area histogram is associated with a different photon number. To describe the observed peaks, we used a model composed of a sum of 16 Gaussian curves.

To assign a detection event with pulse area $a$ to a photon number $m$, we need to determine a set of counting thresholds $t_m$ such that if
\begin{equation}
t_m < a < t_{m+1},
\end{equation}
we assign a photon-number $m$ to this event. These counting thresholds $t_m$ were obtained after normalising each distribution obtained through the area histogram analysis. The counting thresholds $t_m$ are positioned in the intersection point between two adjacent normalised distributions, as shown in the left panel of Figure~\ref{fig:normalised}. Once these $t_m$ are established, assigning a photon-number to a detection event becomes a trivial task, which allows us to perform photon-number-resolving measurements in real time. Note that the distributions are wider for larger areas, thus the uncertainty increases when assigning a photon-number to larger areas. 

To evaluate how precisely we can perform these photon-number assignments based on area measurements, we use the normalised distributions as the underlying distributions of photon numbers $n$ registered by the TES in a given detection event. We are interested in determining $p^n_m$, the probability of correctly assigning the photon-number $m$ to a detection event where $n$ photons were detected. We calculate this probability using 
\begin{equation}
p^n_m = \int_{t_m}^{t_{m+1}}  g_n(a) da,
\end{equation}
where $g_n(a)$ is the normalised distribution associated with the detection of photon-number $n$. In the insets in Figure~\ref{fig:normalised}, we select two numbers ($n {=} 2$ and $n {=} 15$) to highlight the overlap between adjacent distributions. Note that for $n {=} 2$ we obtain a precision of parts-per-billion, $p^2_2 {=} 0.99999999527$,
with the probability of incorrectly assigning a two event to one photon being $p^2_1 {=} 8.5 {\times} 10^{-10} $; and for assigning it to 3 photons being $p^2_3 {=} 3.88 {\times} 10^{-9}$. For higher photon numbers the discrimination probabilities, $p^i_i$, gradually reduce, and for 16 photons, we have $p^{16}_{16} {=} 90\%$ (all values summarised in Table~\ref{tab:probability} in Appendix~\ref{appendix:photon-number_probabilities}). The decrease in discrimination reflects the decreasing counts available: we used a weak coherent pulse such that for $n{\geq}13$ there are a relatively small number of counts.

\section{Detector Tomography Routine}
\label{sec:det_tomo}

Our last step to characterise the performance of our detection system was to perform a quantum detector tomography routine~\cite{lundeen_tomography_2009}. Quantum detector tomography aims to reconstruct the positive operator-valued measure (POVM) that completely describes the measurement apparatus~\cite{lundeen_tomography_2009, feito_measuring_2009}. Each POVM element $E_m$ describes a possible detection outcome $m$. In our case, each $E_m$ describes a measurement result that we associate with a photon-number detection event of $m$ photons. The probability of observing outcome $m$ is given by
\begin{equation}
	\Pr(m) = \mathrm{Tr}(\rho E_m),
	\label{eq:tomography}
\end{equation}
where $\rho$ is the probe state. To perform detector tomography, we need to use a tomographically complete set of input states. For photon-number-resolving measurements, a set of coherent states is tomographically complete~\cite{lundeen_tomography_2009, feito_measuring_2009}. Our analysis used 4 strongly attenuated coherent states $|\alpha\rangle$ with $\alpha$ varying from 1.38 to 2.16 as input states. The values for $\alpha$ were obtained from fitting coherent states to the counts obtained using the calibration method described in Section~\ref{sec:results}. 

The experimental setup employed is the same described in Section~\ref{sec:results}, with a 820~nm pulsed diode laser being used as the light source. For all intensities, subsequent pulses are separated by 100~$\mu$s. For each intensity, we acquired approximately $1.7 \times 10^7$ counts. To measure the number of vacuum detections, we used the signal from the electric pulse generator as a reference. 
We chose a coincidence interval with 600~ns width (from 100~ns to 700~ns, see Figure~\ref{fig:histogram_times} Appendix~\ref{sec:coincidence}) between the generator and TES pulses to incorporate the majority of photons from the laser while minimising accidental background photons. 
A generator pulse without a corresponding TES pulse was considered a vacuum count. 

We note that, since TES measures the energy of the photon wavepacket, it is a phase insensitive detector~\cite{brida_quantum_2012, humphreys_tomography_2015}. Therefore, only the diagonal of a POVM element will have non-zero terms, and only the intensity $|\alpha|^2$ of the coherent states is sufficient to characterise the input states. For a coherent state $|\alpha\rangle$, the probability of observing $n$ photons in a detection event is given by
\begin{equation}
\Pr(n|\alpha) = e^{-|\alpha|^2/2} \frac{|\alpha|^{2n}}{n!}.
\end{equation}

The tomographic routine consists in inverting Equation~\eqref{eq:tomography} using the recorded data for $\Pr(m)$ and the known set of input states $\{\rho_i\}$ to determine the set of POVM elements $\{E_m\}$. To obtain the POVM, we performed a two-norm minimisation using the convex optimisation package for Python (CVXPY)~\cite{diamond_cvxpy_2016, agrawal_rewriting_2018}. The estimated POVM is the solution to:
\begin{equation}
	E_{\rm est}= \arg \min_{E} \big(||\Pr(m) - F E ||^2_2\big),
	\label{eq:convex_min}
\end{equation} 
where $F$ is a matrix where the rows represent different $\alpha$, and the columns represent their projection for different values of $m$ used. 
$E = {E_m}$ for $m = [0, N]$ is the detector POVM, where $N$ is the maximum number of photons the detector can distinguish. 
The POVM element $E_N$ is defined as 
\begin{equation}
    E_N = \mathbb{1} - \sum_{m=0}^{N-1} E_m,
\end{equation}
where $\mathbb{1}$ is the identity matrix. 
$E_N$ accounts for all detection with $N$ or more photons. 
The constraints over the POVM elements $E_m$ are
\begin{equation}
\sum_{m=0}^{N} E_m = \mathbb{1}, \quad \text{and }  \quad E_m \ge 0,
\end{equation}
to ensure the POVM is physically realisable. 
The estimated POVM elements $E_{{\rm est},m}$ are defined by
\begin{equation}
E_{{\rm est},m}= \sum_{k=0}^N \Pr (m|n) |n \rangle \langle n|,
\label{eq:povm_est}
\end{equation}
where $\Pr(m|n)$ is the probability of reporting $m$ photons given that there were $n$ photons at the input. 
For an ideal POVM element, $E_n = |n\rangle \langle n|$, we would have $\Pr(m|n) = \delta_{m,n}$ in Equation \eqref{eq:povm_est}.

\begin{figure}[!t]
    \includegraphics[width=\columnwidth]{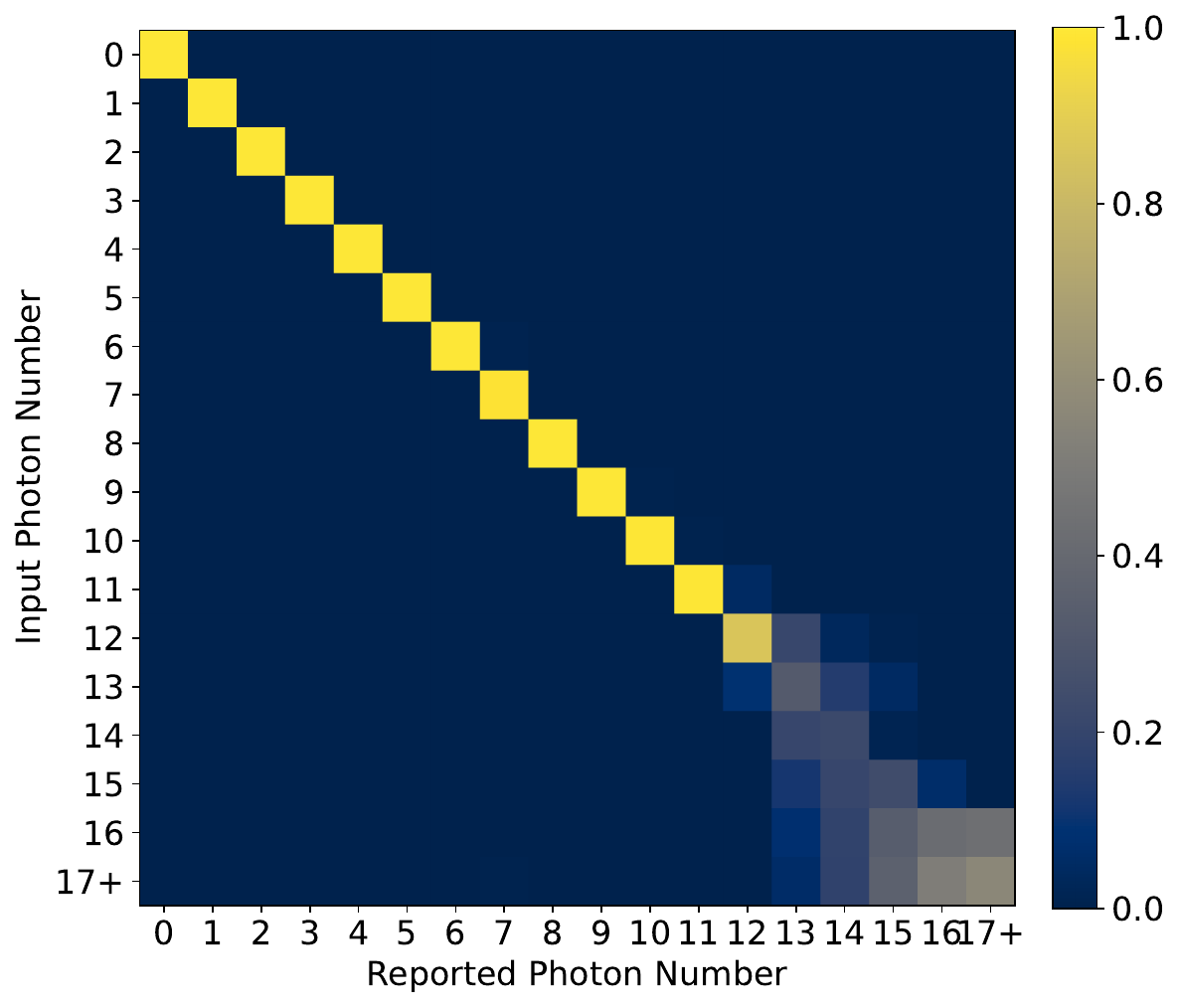}
    \vspace{-7mm}
    \caption{Confusion matrix showing the results for the detector tomography routine. POVM elements are the columns of the confusion matrix, see Equation \eqref{eq:povm_est}. Colour bar at figure right shows the probability of reporting a photon-number $m$ given an input photon-number $n$,  $\Pr(m|n)$. The quality of reconstruction degrades for $n>11$ due to the low intensity of the coherent states used in the tomographic routine. We see high quality reconstruction for $n \leq 11$, with $\Pr(m=n|n) \geq 0.99$ in this range.}
    \label{fig:confusion_matrix}
\end{figure}

We show our results using a confusion matrix, see Figure~\ref{fig:confusion_matrix}, which takes advantage of the fact that all POVM elements are diagonal. Confusion matrices summarise the probabilities of reporting a photon-number $m$ given that you had a detection event with photon-number $n$, i.e., $\Pr( {\rm reported \ } m| {\rm input \ } n)$. In Figure~\ref{fig:confusion_matrix}, each column corresponds to an estimated POVM element $E_{{\rm est},m}$, and each row corresponds to an input photon number $n$. In the ideal case, the confusion matrix would be the identity matrix, with all terms in the diagonal equal to 1, and all non-diagonal terms 0.  We note that the reconstruction quality falls considerably for $n>11$. That is the case even when attempting to reconstruct an ideal POVM with the same input states used in the experimental reconstruction. The small range of intensities used for the input states and the low number of counts for high photon number limit the reconstruction quality in this range. 

The readout fidelity $f_{read}$ provides the average probability of correctly reporting the number of photons in the range analysed, and it is calculated using:
\begin{equation}
f_{read} = \frac{\rm Tr(E)}{N},
\label{eq:fidelity}
\end{equation}
where $N$ is the maximum number of photons in the interval analysed. 
When analysing the range from 0 to 11 or more photons $f_{read} = 0.99$. 
For the full range shown in Figure~\ref{fig:confusion_matrix}, the readout fidelity drops to $f_{read} = 0.80$, a consequence of the low number of counts in the region where $n \geq 11$\footnote{
Due to technical problems with our detectors, we could not take data for coherent states with larger $\alpha$ under the same conditions as the data presented here. 
Nevertheless, we do not expect saturation effects to be relevant within the shown photon-number range. 
A single TES can distinguish photon numbers up to 37 photons, as seen e.g. in Ref.~\cite{eaton_resolution_2023}.}.

Additionally, we investigated the necessary intensity $|\alpha|^2$ of a coherent state to perform quantum detector tomography with readout fidelities {>}{0.99} over the entire photon number range studied here.
We applied our routine in a POVM that describes a detector with 0.01 probability of incorrectly labelling a Fock state $|n\rangle$ over the range between 1 to 16 photons: for example, the detector had 0.005 probability of labelling $|10\rangle$ as $|9\rangle$ or $|11\rangle$, respectively. 
In this scenario, we would need coherent states with $|\alpha|^2$ ranging from 0 up to at least 9 to observe the desired fidelity. 
In a realistic scenario, the probability of correctly assigning a photon number to a Fock state will depend on the Fock state itself and decrease for larger Fock states, further increasing the required $|\alpha|^2$.

\section{Conclusion}

Our FPGA-based hardware processor introduces features desirable across a wide variety of applications. Its parts-per-billion photon-number resolution allows Gottesman-Kitaev-Preskill (GKP) type optical quantum computing, where the detected photon-number output pattern is critical~\cite{bourassa_blueprint_2021}. In addition, number resolution shrinks the logical qubit overheads by minimising error rates, pushing towards thresholds for fault-tolerant quantum computing~\cite{quesada_simulating_2019, su_conversion_2019, tzitrin_progress_2020}, and is highly desirable in Gaussian BosonSampling~\cite{hamilton_gaussian_2017, quesada_gaussian_2018, zhong_quantum_2020}. The reduction in collected and analysed data from the TES will enable new avenues, such as quantum-enhanced imaging where the ability to have even modest arrays of number-resolving detectors allows imaging maximal information extraction at the quantum limit~\cite{howard_optimal_2019}. Possible applications range from astronomy for imaging large, remote bodies~\cite{howard_optimal_2019} to microbiology, where biomolecules can be probed at the single-photon level~\cite{niwa_few-photon_2017}. A single FPGA-hardware processor could handle the readout of a small array, significantly reducing data transfer and storage, as well as the number of data channels. 

Finally, we dedicate this paper to our friend and co-author, the late Sae Woo Nam. We hope we did you proud.

\section*{Acknowledgments}
G.G. and L.A.M. thanks Lewis Howard for useful discussions. L.A.M. acknowledges the support from the Brazilian Agency CAPES (Coordena{\c c}\~ao de Aperfei{\c c}oamento de Pessoal de N\'ivel Superior), Finance Code 88881.128437/2016-01. 
This research was supported by the Australian Research Council Centre of Excellence for Engineered Quantum Systems (EQUS, CE170100009). 

\section*{Disclosures} The authors declare no conflicts of interest. Certain commercial equipment, instruments, or materials are identified in this paper to foster understanding. Such identification does not imply recommendation or endorsement by the National Institute of Standards and Technology, nor does it imply that the materials or equipment identified are necessarily the best available for the purpose.

\newpage
\onecolumn
\appendix
\renewcommand{\thefigure}{A\arabic{figure}}
\renewcommand{\thetable}{A\arabic{table}}
\setcounter{figure}{0}
\setcounter{table}{0}

\section*{Appendix}

\section{Comparison with superconducting nanowire single-photon detectors (SNSPDs)}
\label{appendix:comparison}

Besides TES, SNSPDs are a promising technology for delivering photon-number resolving measurements. One of the main techniques to obtain photon number information with SNSPD is to use a series of detectors in a grid in such a way that the probability of two photons arriving in the same detector is negligible. In Table~\ref{tab:comparison}, we present a comparison of TES with these detectors, taking into account important characteristics, such as: detection efficiency $\eta$, dynamic range, wavelength $\lambda$ range, operating temperature $T_O$, and count rate.

\begin{table*}[!ht]
\vspace{0.5cm}
\centering
\begin{tabular}{lllllll}
Detector Type   & \begin{tabular}[c]{@{}l@{}}Dynamic \\ Range, n \end{tabular} & \begin{tabular}[c]{@{}l@{}}Detection \\ Efficiency\end{tabular} & \begin{tabular}[c]{@{}l@{}}$T_O$ \end{tabular} & \begin{tabular}[c]{@{}l@{}} $\lambda$ range\end{tabular} & \begin{tabular}[c]{@{}l@{}}Count \\ Rate\end{tabular} \\ \hline
TES     & 1 to 16 &    95\%@1550 nm    & 80 mK &    Optical to infra-red & 10 kHz  \\
6-Pixel SNSPD \cite{tao_high_2019}  & 1 to 6  & 68\%@1550 nm & 7.0 K & 1100 to 2000 nm & 300  MHz\\
24-Pixel SNSPD \cite{mattioli_photon-counting_2016} & \begin{tabular}[c]{@{}l@{}}1 to 24  \end{tabular} & 0.8\%@1310 nm  & 1.6 K    & Not specified & 700 MHz  \\
\end{tabular}
\caption{Performance comparison between different types of superconducting photon-detectors with photon-number resolving capabilities.} 
\label{tab:comparison}
\end{table*}

We recently became aware of a very detailed performance comparison in Ref.~\cite{lita_development_2022}, however, detectors are often tailored to excel in one metric and hence the best performance for different figures of merit rarely come from the same citation, as pointed out by the authors.

\section{FPGA Measurements}
\label{appendix:fpga}

\begin{figure}[!htb]
    \centering
    \vspace{-3mm}
    \includegraphics[width=0.8\columnwidth]{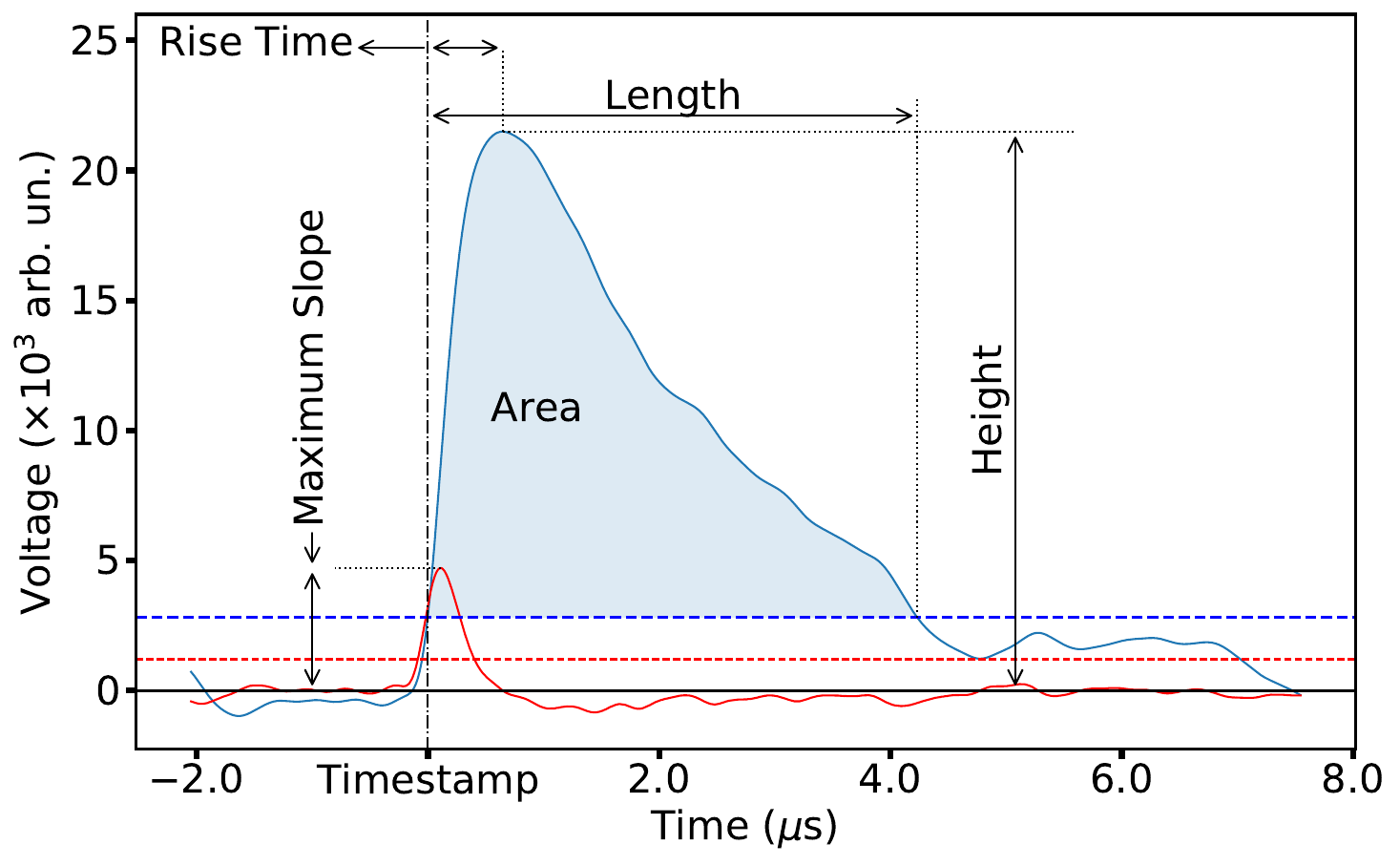}
    \vspace{-3mm} 
    \caption{TES pulse detailed with measurements that our FPGA can perform. The solid blue line is the TES pulse and the solid red line is the TES pulse slope (amplified $1000\times$ for clarity).}
    \label{fig:fpga_with_all}
\end{figure}

The FPGA performing the measurements on the digitised TES signal is a Xilinx 6 series implemented using the ISE toolchain. From the digitised signal, our FPGA derives two signals, as shown in Figure~\ref{fig:fpga_with_all}: the TES signal (solid blue curve) and slope signal (solid red curve). First, the TES signal is derived directly from the ADC output after it passes a digital low-pass filter, which is used to remove high frequency noise present in the TES pulses. The slope signal is derived from the TES signal using a digital filter configured as a differentiator.
\vspace{2mm}

Rises in the data stream are identified using the slope signal: they start when the slope signal becomes positive and finish when it becomes negative. Once a rise has been identified, it needs to satisfy two conditions to be considered a valid detection. First, the maximum slope achieved during a rise must be larger than the slope threshold (dashed red line). Second, the height of the TES signal must be larger than the pulse threshold (dashed blue line). These detection thresholds were introduced to allow the hardware processor to distinguish between rises associated with an actual photon detection from smaller rises due to electronically associated signal fluctuations. Both detection thresholds are chosen by the user based on the information gathered and displayed by the multi-channel analyser, as described in the main text.

We also measure the pulse rise time from the TES pulse, which is the time taken from the moment the timestamp is given until it reaches the maximum height (Figure~\ref{fig:fpga_with_all}). Rise time must be considered when considering appropriate time windows, e.g. for counting coincidences. Additionally, we measure the number of rises in each TES pulse, which can be used to indicate the presence of background counts, for example. Additional rises indicate that two or more absorption events happened before the TES could cool down to its operating temperature, as shown in Figure~\ref{fig:multi_absorption}. Based on this information, the user can perform a more careful analysis to check if it is possible to assign photon number to different events or to discard the pulse, if the information available is insufficient to assign a photon number with reasonable certainty. We note that our current implementation cannot perform height and maximum slope measurements simultaneously.
\vspace{1mm}

\begin{figure}[!b]
    \centering
    \vspace{-3mm}
    \includegraphics[width=0.5\columnwidth]{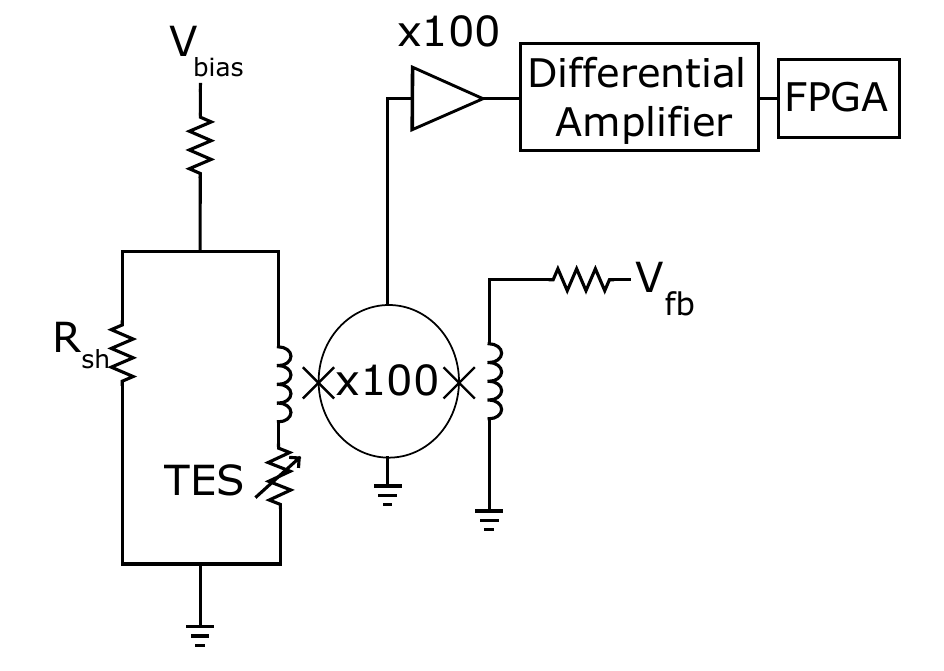}
    \vspace{-3mm} 
    \caption{TES bias and readout system. A stable-room temperature current source and a shunt resistance $R_{sh}$ ($\approx 20~m \Omega$) are used to provide the TES voltage bias. When a photon absorption event occurs, the variation in current is amplified using an array of DC superconducting quantum interference devices (SQUID), which are inductively coupled to the TES circuit. These low-noise amplifiers are maintained at 4~K and amplify the TES signal 100 times. The SQUID output voltage is again amplified 100 times at room temperature and then sent to the differential amplifier. Finally, it passes through an analogue-to-digital converter and is fed to the FPGA, where all time and pulse shape measurements are performed. The voltage feedback, $V_{FB}$, is used to bias the SQUIDS. }
    \label{fig:circuitry}
\end{figure}

\section{TES Detectors}
\label{appendix:tes-detectors}

Our TES detectors were fabricated at the National Institute of Standards and Technology (NIST) Boulder and are thin films of tungsten, with a surface area of 25~$\mu$m\textsuperscript{2}, thickness of 20~nm, fabricated on silicon. NIST reported efficiency of at least 95\% for these detectors~\cite{lita_counting_2008}; in-situ efficiency measurements were not performed. The tungsten thin films are embedded in an optical structure to optimise the absorption at 820~nm wavelength. We use an adiabatic demagnetisation refrigerator (ADR) to maintain the detectors at an operating temperature of 80~mK, below the critical temperature of the tungsten film. The detectors are voltage biased, and after a detection event occurs their initial temperature is restored due to negative electro-thermal feedback. The detection circuitry is displayed in Figure~\ref{fig:circuitry}.
\vspace{2mm}

\section{Multi-absorption events}
\label{appendix:multi-events}

\begin{figure}[!htb]
    \centering
    \vspace{-3mm}
    \includegraphics[width=0.7\columnwidth]{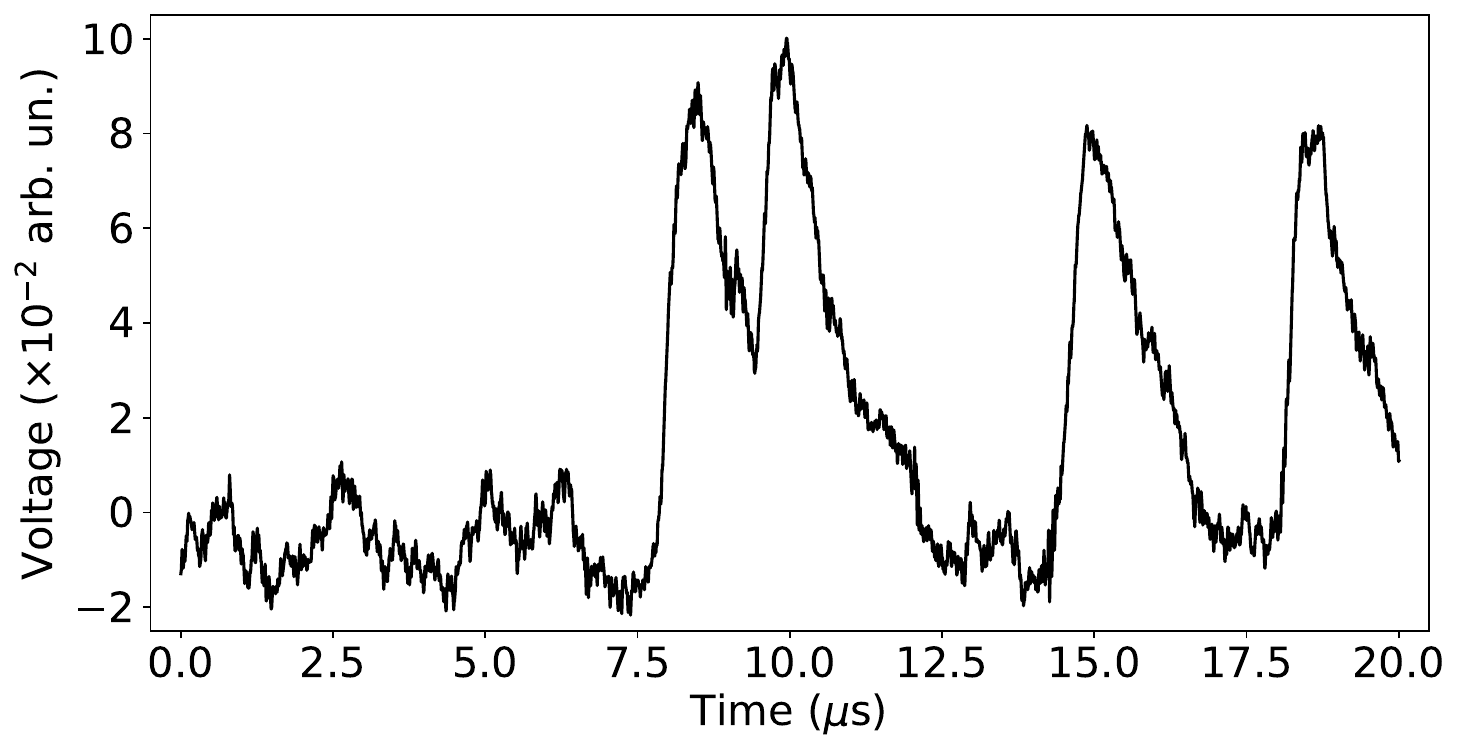}
    \vspace{-3mm}
    \caption{TES analogue output showing a multi-absorption event obtained using an oscilloscope. Our FPGA can identify the multiple rises in a detection event and measure the height of each rise. The user can choose to develop a method to analyse these cases or to remove the events where multi-absorption events occur.}
    \label{fig:multi_absorption}
\end{figure}

TES have no dead time, and events with different photons being absorbed at different times can be recorded, as shown in Figure~\ref{fig:multi_absorption}. A second detection event occurs at around 10~$\mu$s, before the detector had enough time to cool down after the first detection. Our FPGA can identify such events by identifying multiple rises in a single detection event and storing the height for different rises. Since multi-absorption events add an extra layer of complexity when performing the photon-number assignment task, in the main paper we used a pulsed light source much slower (10~kHz) than the longest cool down times ($\approx$~10~$\mu$s) to avoid multi-absorption events.
\vspace{0.5cm}

\section{Area Histogram}
\label{appendix:area_histogram}

\begin{figure}[!t]
    \centering
    \includegraphics[width=0.49\columnwidth]{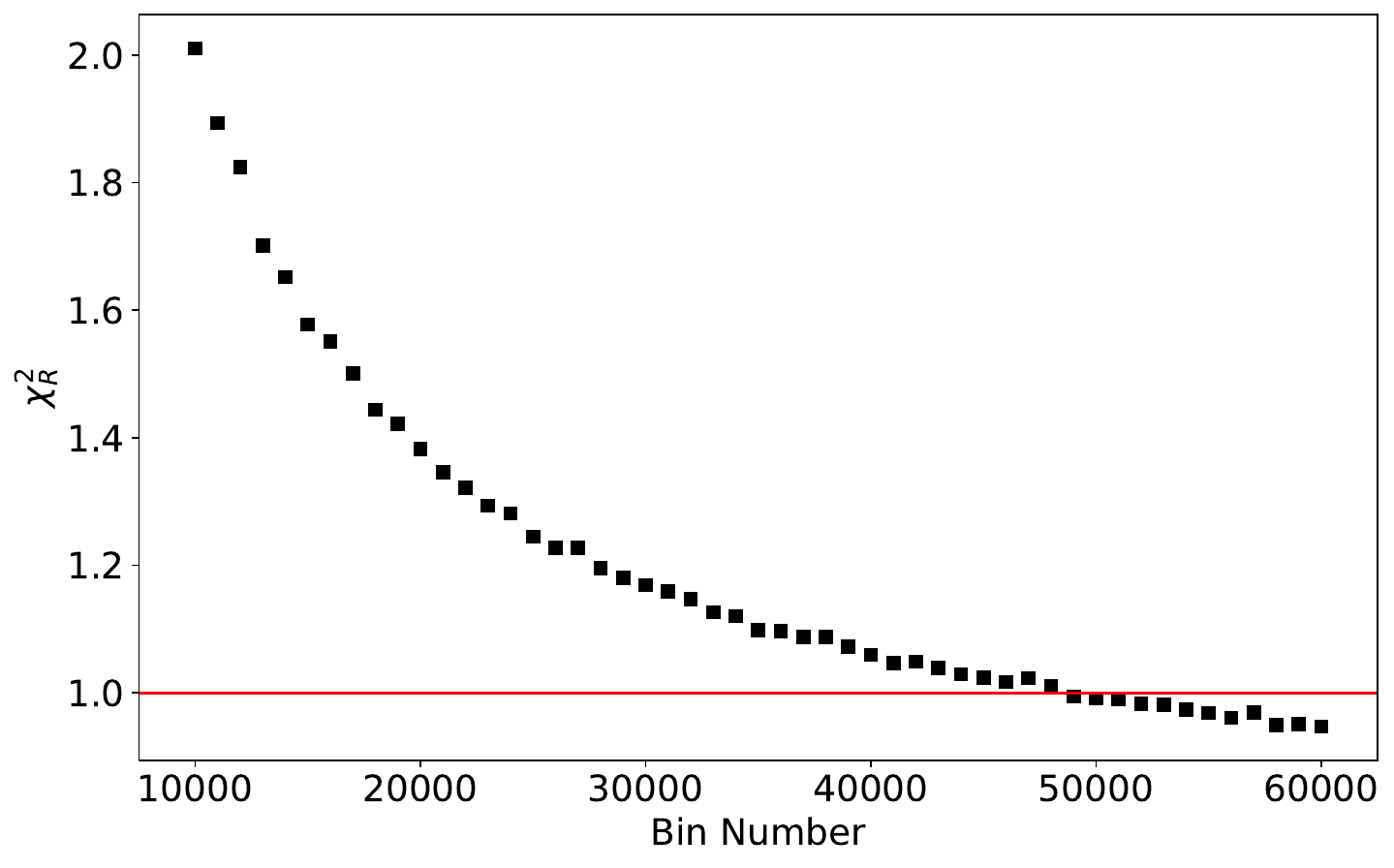}
    \includegraphics[width=0.49\columnwidth]{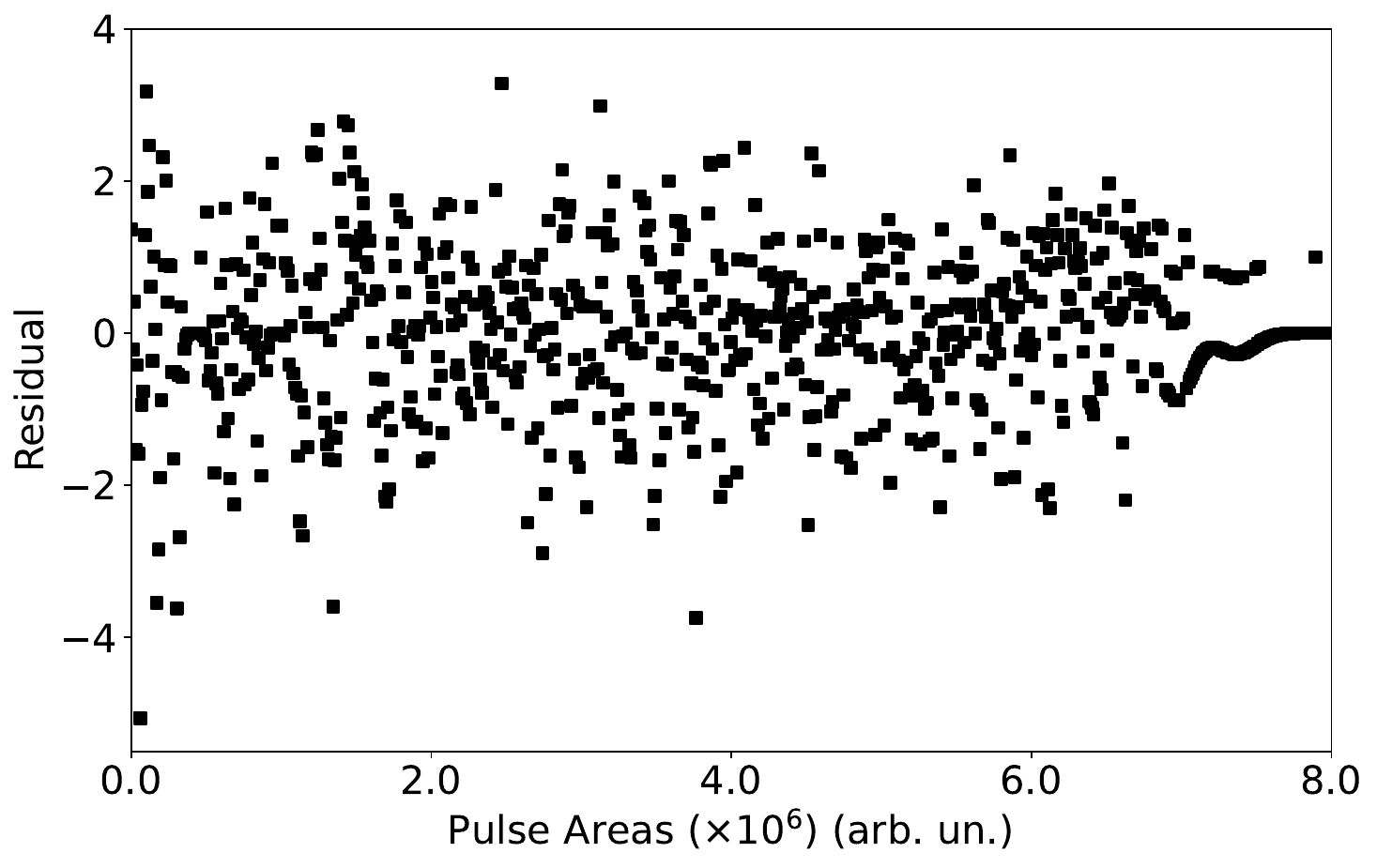}
    \vspace{-3mm}
    \caption{(\textit{Left}) $\chi_R^2$ for our model $M$  as a function of the number of bins used in the area histogram. (\textit{Right}) Residual calculated from Equation \eqref{eq:residual}. }
    \label{fig:redchi}
\end{figure}

The model $M$ we used to fit the area histogram---the dashed red line in Figure~\ref{fig:histogram_area}---is composed of a sum of Gaussian curves $G_n(a)$, defined by
\begin{equation}
    M(A_1, \mu_1, \sigma_1, ..., A_{16}, \mu_{16}, \sigma_{16}) = \sum_{n=1}^{16} G_n(a),
\end{equation}
where
\begin{equation}
    G_n(a) = \frac{A_n}{\sqrt{2\pi} \sigma_n} \exp \bigg( \frac{-(a - \mu_n)^2}{2 \sigma_n^2}\bigg),
\end{equation}
with $A_n$ being the amplitude, $\mu_n$ the mean, and  
$\sigma_n$ is the standard deviation of the Gaussian used to describe peak $n$. 
To evaluate the goodness of the fit, we used the reduced chi-squared, defined as 
\begin{equation}
    \chi_R^2 = \sum_i \frac{(C_i - M)^2}{k \epsilon_i^2}, 
\end{equation}
where $C_i$ is the number of counts in the i$^{\rm th}$ bin, $\epsilon_i$ is the error in the i$^{\rm th}$ bin, and $k$ is the number of degrees of freedom. We define $k$ as
\begin{equation}
    k = n_{bins} - n_{par},
\end{equation}
where $n_{bins}$ is the number of bins in the histogram and $n_{par}$ is the number of parameters in the fit. To determine $n_{bin}$ we observed how the $\chi_R^2$ varied as function of the number of bins used, see left panel of Figure~\ref{fig:redchi}. To calculate $\chi_R^2$, we assumed our data follows Poissonian statistics, therefore $\epsilon_i = \sqrt{C_i}$.  We divide the analysed area interval of $[0, 10 \times 10^6 ]$ into 45000 bins, as we obtained $\chi_R^2 \approx 1$ in this case. The number of detection events used was $\approx 1.7 \times 10^7$. The residual $Res$, shown in the right panel of Figure \ref{fig:redchi}, was calculated using 
\begin{equation}
    \label{eq:residual}
    Res_i {=} \frac{C_i {-} M}{\sqrt{C_i}}.
\end{equation}

To determine the counting threshold position $t_m$, we used the normalised distributions $g_n(a)$ given by
\begin{equation}
    g_n(a) = \frac{1}{\sqrt{2\pi} \sigma_n} \exp \bigg( \frac{-(a - \mu_n)^2}{2 \sigma_n^2}\bigg),
\end{equation}
which has an area equal to 1.

\begin{figure}[!ht]
    \centering
    \vspace{-3mm}
    \includegraphics[width=0.8\columnwidth]{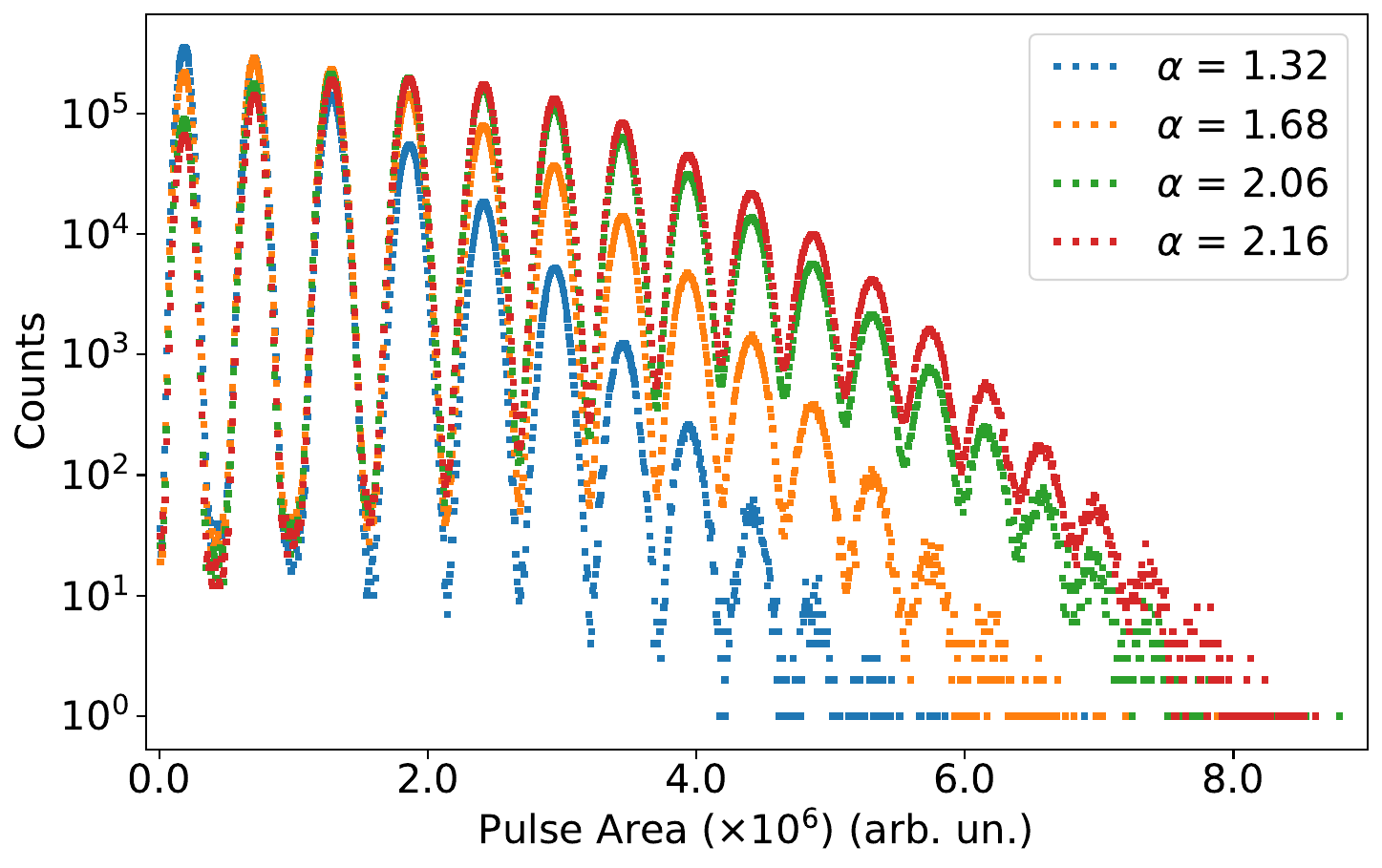}
    \vspace{-3mm} 
    \caption{Area histogram for different intensities registered with our detection system. Each colour corresponds to a different intensity of the light source. Note that the peaks---which correspond to a given photon number--- occur in the same position independent of the intensity used. }
    \label{fig:intensity_diff}
\end{figure}

Once the counting thresholds $t_m$ were established, we calculated the $\chi_R^2$ for each $g_n(a)$ to quantify how well a single Gaussian could describe the data points inside the area interval $[t_m, t_{m+1}]$, see Table~\ref{tab:chi}. Note that for high photon numbers ($n>14$), the $\chi_R^2$ is smaller than 1. This is an expected result, since this is a low count region due to the relatively low intensity of the light source used. In Figure~\ref{fig:intensity_diff} we present histograms for different intensities, showing that the source's intensity can limit the highest number of photons we can distinguish using this calibration process.

\begin{table}[!b]
\centering
\begin{tabular}{cccccccccccccc}
Peak & $\chi_R^2$ &  &  & Peak & $\chi_R^2$ &  &  & Peak & $\chi_R^2$ &  &  & Peak & $\chi_R^2$ \\ \cline{1-2} \cline{5-6} \cline{9-10} \cline{13-14} 
1    & 2.56       &  &  & 5    & 1.06       &  &  & 9    & 1.27       &  &  & 13   & 1.20       \\
2    & 0.98       &  &  & 6    & 1.19       &  &  & 10   & 1.20       &  &  & 14   & 0.96       \\
3    & 2.72       &  &  & 7    & 1.65       &  &  & 11   & 1.21       &  &  & 15   & 0.47       \\
4    & 1.72       &  &  & 8    & 1.65       &  &  & 12   & 1.18       &  &  & 16   & 0.20   
\end{tabular}
\caption{Reduced chi-squared, $\chi_R^2$ , between model and data for all peaks. They were calculated by considering the Gaussian distribution $g_n(a)$ used to describe a peak in the interval $[t_n, t_{n+1}]$. }
\label{tab:chi}
\end{table}

To show that this calibration procedure is robust within the intensity range, we estimated $\alpha$ for 4 different coherent states using different calibrations. Besides the calibration reported in the main paper, we also calibrated the detector using the coherent state with $\alpha = 1.32$ (blue dots in Figure~\ref{fig:intensity_diff}). In this case, our model was composed of 10 Gaussian curves due to the lower intensity of the coherent state used for calibration. The counting thresholds $t_n$ were obtained using the same method described before. In Table~\ref{tab:intensity_diff}, we compare the estimated values for $\alpha$ for 4 different coherent states using both calibrations. For the left and right tables, the values of $\alpha$ for the coherent state used for calibration were $1.32$ and $2.16$, respectively. Note that the $\alpha$ values for the entire set agree within one standard deviation, showing that our calibration provides consistent estimation for the coherent states used in this low intensity range. We highlight that the standard deviation presented here is the value provided by the fitting algorithm and does not represent the experimental uncertainty with the coherent state used for calibration.  

\begin{table}[!hb]
\centering
\begin{tabular}{c|c|c}
$\alpha$ & $\sigma$     & $\chi^2_{red}$ \\ \hline
1.3245   & $2\times10^{-4}$  & 1.18         \\
1.6880   & $2 \times 10^{-4}$  & 1.18         \\
2.0627   & $2 \times 10^{-4}$  & 0.88         \\
2.1648   & $2 \times 10^{-4}$  & 3.88        
\end{tabular}
\qquad
\begin{tabular}{c|c|c}
$\alpha$ & $\sigma$          & $\chi^2_{red}$ \\ \hline
1.3245   & $2 \times 10^{-4}$ & 1.22         \\
1.6880   & $2 \times 10^{-4}$ & 1.24         \\
2.0627   & $2 \times 10^{-4}$ & 1.81         \\
2.1649   & $3 \times 10^{-4}$ & 5.33        
\end{tabular}

\caption{Comparison between estimated values for $\alpha$ using two different coherent states for calibration. For the left table, the coherent state used for calibration had $\alpha$ = 1.32. For the right table, $\alpha$ = 2.16. We note that all values agree within one standard deviation $\sigma$.}
\label{tab:intensity_diff}
\end{table}

\section{Photon-Number Assignment Probabilities}
\label{appendix:photon-number_probabilities}

The discrimination problem: once a photon detection event has been recorded, we want to determine the number of photons in that event. Table~\ref{tab:probability} lists the probabilities of correctly, $m{=}n$, or incorrectly, $m{\neq}n$, assigning a photon number $m$ to a detection event with $n$ photons, $p^n_m$. Since only the cases where $m{=}n{-}1$ and $m{=}n{+}1$ have non-negligible values, we present only these three columns. Note that for 17 photons or more, we can no longer discriminate the photon number, therefore $p^{n}_{17+}$ should be read as the probability of assigning 17 photons \emph{or more} to a detection event where $n$ photons were detected. The uncertainties in the probabilities of making a correct assignment $p^n_{m{=}n}$ were calculated by considering the uncertainties in the position of the detection thresholds $t_m$. Code available on request.
\vspace{1mm}

\begin{table}[!ht]
\centering
\begin{tabular}{@{}cccc@{}}
n  & $p^n_{m=n-1}$               & $p^n_{m=n}$        & $p^n_{m=n+1}$ \\ \midrule
1  & $ 4.01190 \times 10^{-6}  $ & $ 0.99999598734^{(2)}_{(2)} $  & $ 7.5  \times 10^{-10} $  \\
2  & $ 8.5     \times 10^{-10} $ & $ 0.99999999527^{(5)}_{(5)} $  & $ 3.88 \times 10^{-9} $  \\
3  & $ 5       \times 10^{-9}  $ & $ 0.999999666^{(3)}_{(4)} $    & $ 3.31 \times 10^{-7} $  \\
4  & $ 3.8     \times 10^{-7}  $ & $ 0.99999343^{(4)}_{(4)} $     & $ 6.19 \times 10^{-6} $  \\
5  & $ 6.6     \times 10^{-6}  $ & $ 0.9999625^{(2)}_{(2)} $      & $ 3.09 \times 10^{-5} $  \\
6  & $ 3.19    \times 10^{-5}  $ & $ 0.9998538^{(5)}_{(5)} $      & $ 1.142 \times 10^{-4} $ \\
7  & $ 1.21    \times 10^{-4}  $ & $ 0.999430^{(2)}_{(2)} $       & $ 4.48 \times 10^{-4} $  \\
8  & $ 4.89    \times 10^{-4}  $ & $ 0.998007^{(8)}_{(9)} $       & $ 1.50 \times 10^{-4} $  \\
9  & $ 1.63    \times 10^{-3}  $ & $ 0.99508^{(2)}_{(2)} $        & $ 3.28 \times 10^{-3} $  \\
10 & $ 3.41    \times 10^{-3}  $ & $ 0.99155^{(3)}_{(4)} $        & $ 5.03 \times 10^{-3} $  \\
11 & $ 5.08    \times 10^{-3}  $ & $ 0.98840^{(6)}_{(7)} $        & $ 6.52 \times 10^{-3} $  \\
12 & $ 6.6     \times 10^{-3}  $ & $ 0.9858^{(1)}_{(1)} $         & $ 7.6 \times 10^{-3} $   \\
13 & $ 7.5     \times 10^{-3}  $ & $ 0.9789^{(7)}_{(7)} $         & $ 1.36 \times 10^{-2} $  \\
14 & $ 1.7     \times 10^{-2}  $ & $ 0.957^{(2)}_{(3)} $          & $ 2.6 \times 10^{-2} $   \\
15 & $ 2.6     \times 10^{-2}  $ & $ 0.936^{(4)}_{(7)} $          & $ 3.7 \times 10^{-2} $   \\
16 & $ 5       \times 10^{-2}  $ & $ 0.90^{(2)}_{(2)} $           & $ 5 \times 10^{-2} $     \\ 
\end{tabular}
\caption{Probabilities of assigning a photon number $m$ to a detection event with $n$ photons. The middle column is the probability of a correct assignment, $m{=}n$; the left and right columns are, respectively, probabilities of the incorrect assignments $m{=}n{-}1$ and $m{=}n{+}1$. The upper and lower numbers in parenthesis indicate the error (one standard deviation) in the reported values for correct probability assignments. \vspace{-5mm}}
\label{tab:probability}
\end{table}

\section{Coincidence window for tomography}
\label{sec:coincidence}

\begin{figure}[!ht]
    \centering
    \vspace{-3mm}
    \includegraphics[width=0.6\columnwidth]{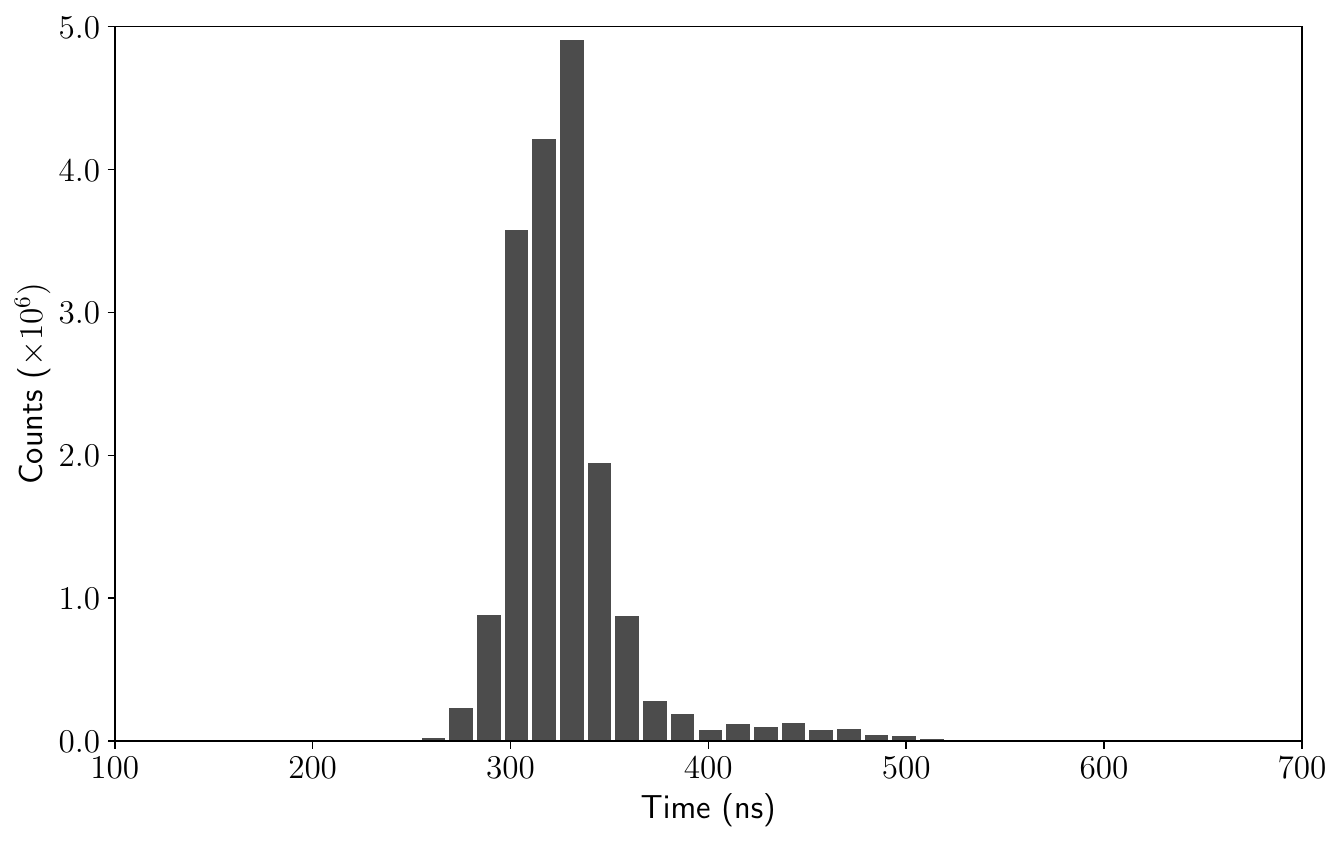}
    \vspace{-3mm} 
    \caption{Histogram of relative times of detection events in a 600~ns window with respect to the registration of an initial electric pulse from the laser.}
    \label{fig:histogram_times}
\end{figure}

As stated in the main paper, we set a coincidence window of 600~ns for the tomography analysis. The window for the detection events to be considered valid spanned from 100~ns to 700~ns after a detection pulse was registered. In Figure~\ref{fig:histogram_times}, we show the relative time the detection events were reported after an electric pulse was registered in the FPGA. We note that 525 detection events were outside this time range, out of approximately $1.8 \times 10^7$ counts. From these, 518 were 1-photon detection events, most likely stray thermal background photons.

\newpage
    
\bibliography{tes_paper}{}
\bibliographystyle{elsarticle-num}

\end{document}